\documentclass[aps,pra,numerical,preprint,showpacs]{revtex4-1}  
\usepackage{graphicx}
\usepackage{bm}
\usepackage{amsmath}
\usepackage{amsfonts}
\usepackage[normalem]{ulem} 
\usepackage[colorlinks=true,linkcolor=blue,citecolor=blue ,urlcolor=blue]{hyperref}

\begin{document}


\title{Entropy of entanglement between quantum phases of a three-level matter-radiation
interaction model}


\author{Luis Fernando Quezada}

\author{Eduardo Nahmad-Achar}
\email{nahmad@nucleares.unam.mx}
\affiliation{%
Instituto de Ciencias Nucleares, Universidad Nacional Aut\'onoma de M\'exico, Apartado Postal 70-543, 04510 M\'exico Cd. Mx.}




\begin{abstract}
	We show that the entropy of entanglement is sensitive to the coherent quantum phase transition between normal and super-radiant regions of a system of a finite number of three-level atoms interacting in a dipolar approximation with a one-mode electromagnetic field. The atoms are treated as semi-distinguishable using different cooperation numbers and representations of SU(3), variables which are relevant to the sensitivity of the entropy with the transition. The results are computed for all three possible configurations ($\Xi$, $\Lambda$ and $V$) of the three-level atoms.
\end{abstract}

\keywords{matter-radiation interaction; cooperation number; entropy of entanglement; quantum phase transition; representation theory; residual entropy}


\maketitle


\section{Introduction}
The interaction of two-level identical atoms with a quantised electromagnetic field, using a dipolar approximation, is described by the Dicke Model \cite{key-1}. A particularly interesting phenomenon regarding this and other quantum systems are quantum phase transitions (QPTs), which can be thought of as sudden, drastic changes in the physical properties of the ground state of a system at zero temperature due to the variation of some parameter involved in the modelling Hamiltonian. In 1973, Hepp and Lieb \cite{key-2, key-3}, and Wang and Hioe \cite{key-4} first theoretically proved the existence of a QPT in the Dicke model. To date, this quantum phase transition has been experimentally observed in a Bose-Einstein Condensate coupled to an optical cavity \cite{key-5, key-6} and it has been shown to be relevant to quantum information and quantum computing \cite{key-7, key-8}. Entanglement between the atoms and the field in the Dicke model has also been studied \cite{key-9, key-10}, allowing the identification of both quantum and semi-classical, many-body features.

Generalisations of the Dicke model which consider atoms of three or more levels have been extensively studied \cite{key-11, key-12, key-13, key-14, key-15, key-16, key-17, key-18, key-19, key-20}. These models allow meaningful interactions with two or more modes of the electromagnetic field, a feature that has been exploited for the development of certain types of quantum memories \cite{key-21, key-22, key-23, key-24}.

An important aspect of these matter-radiation interaction models is the distinguishability of the atoms, a characteristic that depends on the space we choose for the Hamiltonian to act on. Most works on the subject treat the atoms as completely indistinguishable; nevertheless, this may not correctly describe some of the experimental realizations of the models. In order to gain distinguishability we must add information of the atomic field to the states we use to describe it, and one possible information we can add is the  \textit{cooperation number}.

The term ``cooperation number'' was first introduced by Dicke in his original paper \cite{key-1}, referring to the different representations of $SU(2)$ used in the description of the full state's space of his Hamiltonian, and whose physical interpretation is that of an effective number of atoms in the system, i.e. the number of atoms that contribute to the energy of the atomic field. The influence of the cooperation number over the QPT, expectation values and entropy of entanglement has already been studied for two-level systems \cite{key-25}.

In this work we study the correlation between the entropy of entanglement and the coherent quantum phases of a system of a finite number of three-level atoms interacting in a dipolar approximation with a one-mode electromagnetic field. Here, using different cooperation numbers and representations of SU(3), we are able to treat the atoms as semi-distinguishable. This correlation by itself suggests the existence of quantum phases for a finite number of semi-distinguishable atoms and has a direct relation with the residual entropy of the system, as the number of possible states at zero temperature would be greater than one.

\section{Theoretical Framework}

\subsection{Modelling Hamiltonian}
The Hamiltonian describing the interaction, in a dipolar approximation, between N three-level identical atoms (same energy levels) and one-mode of an electromagnetic field in an ideal cavity, has the expression ($\hbar=1$) \cite{key-18}

\begin{equation}
H=\overline{\omega}_{1}e_{11}+\overline{\omega}_{2}e_{22}+\overline{\omega}_{3}e_{33}+\Omega a^{\dagger}a-\frac{1}{\sqrt{N}}\sum_{i<j}^{3}\mu_{ij}\left(e_{ij}+e_{ij}^{\dagger}\right)\left(a+a^{\dagger}\right).\label{eq:1}
\end{equation}

Here, $\overline{\omega}_{1}$, $\overline{\omega}_{2}$ and $\overline{\omega}_{3}$ are the three energy levels of the atoms, with $\overline{\omega}_{1} \leq \overline{\omega}_{2} \leq \overline{\omega}_{3}$, $\Omega$ is the frequency of the field's mode, $\mu_{ij}$ are the dipolar coupling parameters between levels $i$ and $j$, $a$ and $a^{\dagger}$ are the annihilation and creation operators of the harmonic oscillator and $e_{ij}$ are the collective atomic matrices, i.e. summations (with as many summands as atoms in the system) of the single-entry matrices $\left(\overline{e}_{ij}\right)_{mn}=\delta_{im}\delta_{jn}$. Choosing the zero of the energy to be at $\frac{1}{3}\left(\overline{\omega}_{1}+\overline{\omega}_{2}+\overline{\omega}_{3}\right)$ we can rewrite this hamiltonian (\ref{eq:1}) in the more useful form

\begin{equation}
H=\omega_{1}J_{z}^{\left(1\right)}+\omega_{2}J_{z}^{\left(2\right)}+\Omega a^{\dagger}a-\frac{1}{\sqrt{N}}\sum_{i<j}^{3}\mu_{ij}\left(e_{ij}+e_{ij}^{\dagger}\right)\left(a+a^{\dagger}\right),\label{eq:2}
\end{equation}
where $\omega_{1}=-\frac{4}{3}\overline{\omega}_{1}+\frac{2}{3}\overline{\omega}_{2}+\frac{2}{3}\overline{\omega}_{3}$, $\omega_{2}=-\frac{2}{3}\overline{\omega}_{1}-\frac{2}{3}\overline{\omega}_{2}+\frac{4}{3}\overline{\omega}_{3}$, $J_{z}^{\left(1\right)}=\frac{1}{2}\left(e_{22}-e_{11}\right)$ (half the population difference between the second and first levels) and $J_{z}^{\left(2\right)}=\frac{1}{2}\left(e_{33}-e_{22}\right)$ (half the population difference between the third and second levels).

\begin{figure}
\centering
\includegraphics[scale=0.7]{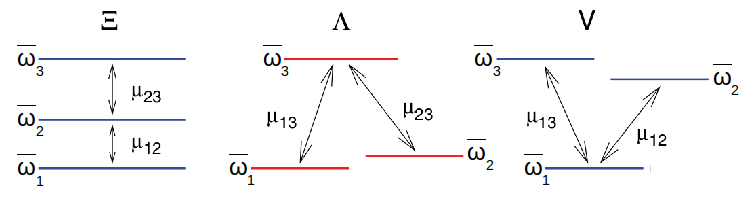}
\caption{Diagram showing the three possible configurations of a three-level atom according to the permitted transitions between its levels.}\label{fig:1}
\end{figure}

Due to selection rules, the parity of the quantum states between which a dipolar transition is made, must be opposite. This forces one of the coupling parameters $\mu_{ij}$ to be zero, giving rise to three possible three-level atom configurations: $\Xi$ configuration ($\mu_{13}=0$), $\Lambda$ configuration ($\mu_{12}=0$) and $V$ configuration ($\mu_{23}=0$) (figure \ref{fig:1}). In this work we consider all three of them.

\subsection{Representation theory and cooperation number}

The operators $J_{z}^{\left(1\right)}$, $J_{z}^{\left(2\right)}$, $e_{12}$, $e_{23}$, $e_{12}^{\dagger}$ and $e_{23}^{\dagger}$ in Hamiltonian (\ref{eq:2}) form a basis for the Lie Algebra of $SU(3)$, thus it is natural to think that its representation theory can provide some insights into the understanding of the modelled system. In fact, this basis has a feature that makes it particularly convenient if one also adopts the labelling scheme for the basis states of the irreducible representations (irreps) of $SU(n)$ devised by Gelfand and Tsetlin \cite{key-29}: these basis states are simultaneous eigenstates of the operators $J_{z}^{\left(1\right)}$ and $J_{z}^{\left(2\right)}$, and explicit formulae exist for the matrix elements of $e_{12}$, $e_{23}$, $e_{12}^{\dagger}$ and $e_{23}^{\dagger}$. In a nutshell, the labelling scheme for the basis states of a given irrep $h=(h_{1},h_{2},h_{3})$ of $SU(3)$, called a Gelfand-Tsetlin pattern, is as follows:

\begin{center}
$\left|
	\begin{array}{ccc}
h_{1} & h_{2} & h_{3}\\
q_{1} & q_{2}\\
r
	\end{array}
\right\rangle$
\end{center}
where the top row contains the information that specifies the irrep, while the entries of lower rows are subject to the {\it betweenness conditions}: 
$h_{1}\geq q_{1}\geq h_{2}$, 
$h_{2}\geq q_{2}\geq h_{3}$ and 
$q_{1}\geq r\geq q_{2}$.

Using these basis states to describe the matter subsystem of our Hamiltonian allows us to have a very simple physical interpretation of the parameters in the Gelfand-Tsetlin pattern: $r$ is the number of atoms in the first (lowest) energy level, $q_{1}+q_{2}-r$ is equal to the number of atoms in the second energy level and $h_{1}+h_{2}+h_{3}-q_{1}-q_{2}$ is equal to the number of atoms in the third (highest) energy level, where $h_{1}$, $h_{2}$ and $h_{3}$ are subject to the constraint $h_{1}+h_{2}+h_{3}=N$ (the total number of atoms). The cooperation number in this description is $h_1 - h_3$.

Representation theory allows us to decompose the space of states (of the matter subsystem) into a direct sum of subspaces labelled by the parameters $h_{1}$, $h_{2}$ and $h_{3}$ (the permitted representations for a given $N$), each representation may appear more than once in the decomposition, the number of times it appears is called the representation's multiplicity. If we were to consider every possible representation with its own multiplicity, we would be treating the atoms as fully distinguishable, on the other hand, if we just consider the symmetric representation ($h_{1}=N$, $h_{2}=h_{3}=0$), we would be treating the atoms as fully indistinguishable. In this work we consider every possible representation but ignore its multiplicity, leading us to treat the atoms as semi-distinguishable, the cooperation number being what adds some distinguishability to the states.

Coherent states of $SU(3)$ are defined as

\begin{equation}\left|\bar{\gamma},\bar{h}\right\rangle_{NN} :=e^{\gamma_{3}e_{12}^{\dagger}+\gamma_{2}e_{13}^{\dagger}+\gamma_{1}e_{23}^{\dagger}}
 \left|\begin{array}{ccc}
h_{1} & h_{2} & h_{3}\\
h_{1} & h_{2}\\
h_{1}
\end{array}\right\rangle\end{equation}
and we take the tensor product of these with the usual coherent states for the harmonic oscillator for the field, as our trial states for a variational procedure, where, following the catastrophe formalism, the expectation value of the Hamiltonian with respect to these trial states is minimised in order to find the critical points and the ground state of the system~\cite{key-30}. As our system is not integrable, and the expression for the expectation value of $H$ is unwieldy, this minimisation is carried out numerically.

\subsection{Entropy of entanglement ($S_{\varepsilon}$)}

Entropy of entanglement is defined for a bipartite system as the von Neumann entropy of either of its reduced states, that is, if $\rho$ is the density matrix of a system in a Hilbert space $\mathcal{H}=\mathcal{H}_{1}\otimes\mathcal{H}_{2}$, its entropy of entanglement is defined as

\begin{equation}
S_{\varepsilon}:=-Tr\left\{ \rho_{1}\log\rho_{1}\right\} =-Tr\left\{ \rho_{2}\log\rho_{2}\right\} ,
\end{equation}

where $\rho_{1}=Tr_{2}\left\{ \rho\right\} $ and $\rho_{2}=Tr_{1}\left\{ \rho\right\}$.

Our Hamiltonian (\ref{eq:2}) models a bipartite system formed by matter and radiation subsystems, which means that the entropy of entanglement can give some insight on the study of the quantum phases; this we analyse below.

\subsection{Fidelity between neighbouring states (${F}$)}

Fidelity is a measure of the ``distance'' between two quantum states; given $\left|\phi\right\rangle$ and $\left|\varphi\right\rangle $ it is defined as

\begin{equation}
F(\phi,\varphi):=\left|\left\langle \phi|\varphi\right\rangle \right|^{2}.
\end{equation}

Across a QPT the ground state of a system suffers a sudden, drastic change, thus it is natural to expect a drop in the fidelity between neighbouring states near the transition. This drop has been, in fact,
already shown to happen \cite{key-26,key-27}. In this work we use the drop in the fidelity between neighbouring coherent states as a characterization of the QPT in the thermodynamic limit.

 
\section{Results}

In this work we studied a system, described by the Hamiltonian (\ref{eq:2}), of four three-level atoms interacting with a one-mode electromagnetic field, thus we had four possible representations (and cooperation numbers) of SU(3), namely $h=(4,0,0)$ (the symmetric one), $h=(3,1,0)$, $h=(2,2,0)$ and $h=(2,1,1)$ with a cooperation number of 4, 3, 2 and 1 respectively. We compared the entropy of entanglement to the fidelity between neighbouring coherent states as functions of the coupling parameters $\mu_{ij}$. Here, based in the results obtained for two-level systems \cite{key-25}, we expected to see a correlation between the coherent quantum phase transition (characterized by the the drop in the fidelity) and the region where the entropy of entanglement reaches its highest values.


\begin{figure}
\centering
\includegraphics[scale=0.29]{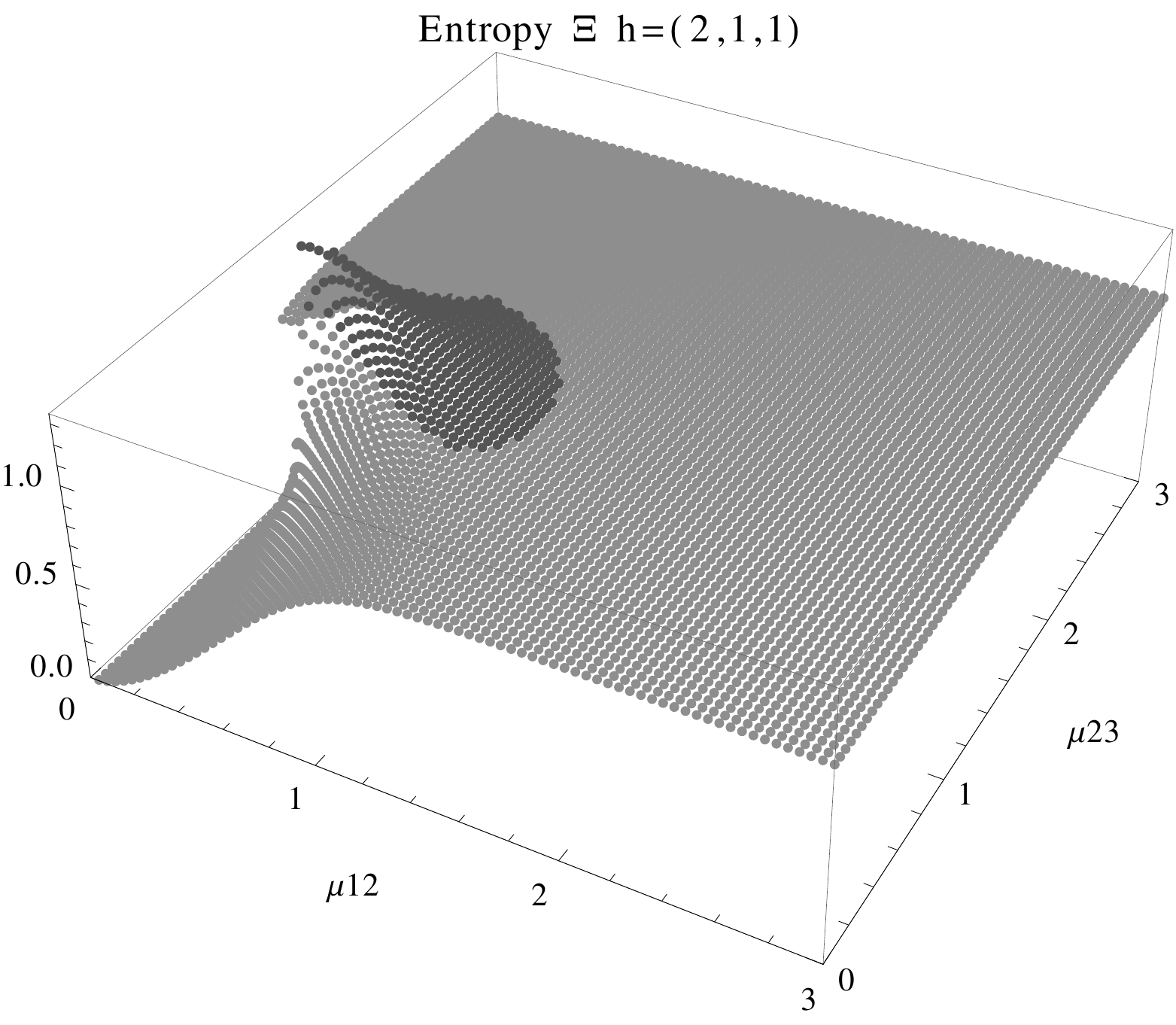}\quad{}\quad{}\includegraphics[scale=0.29]{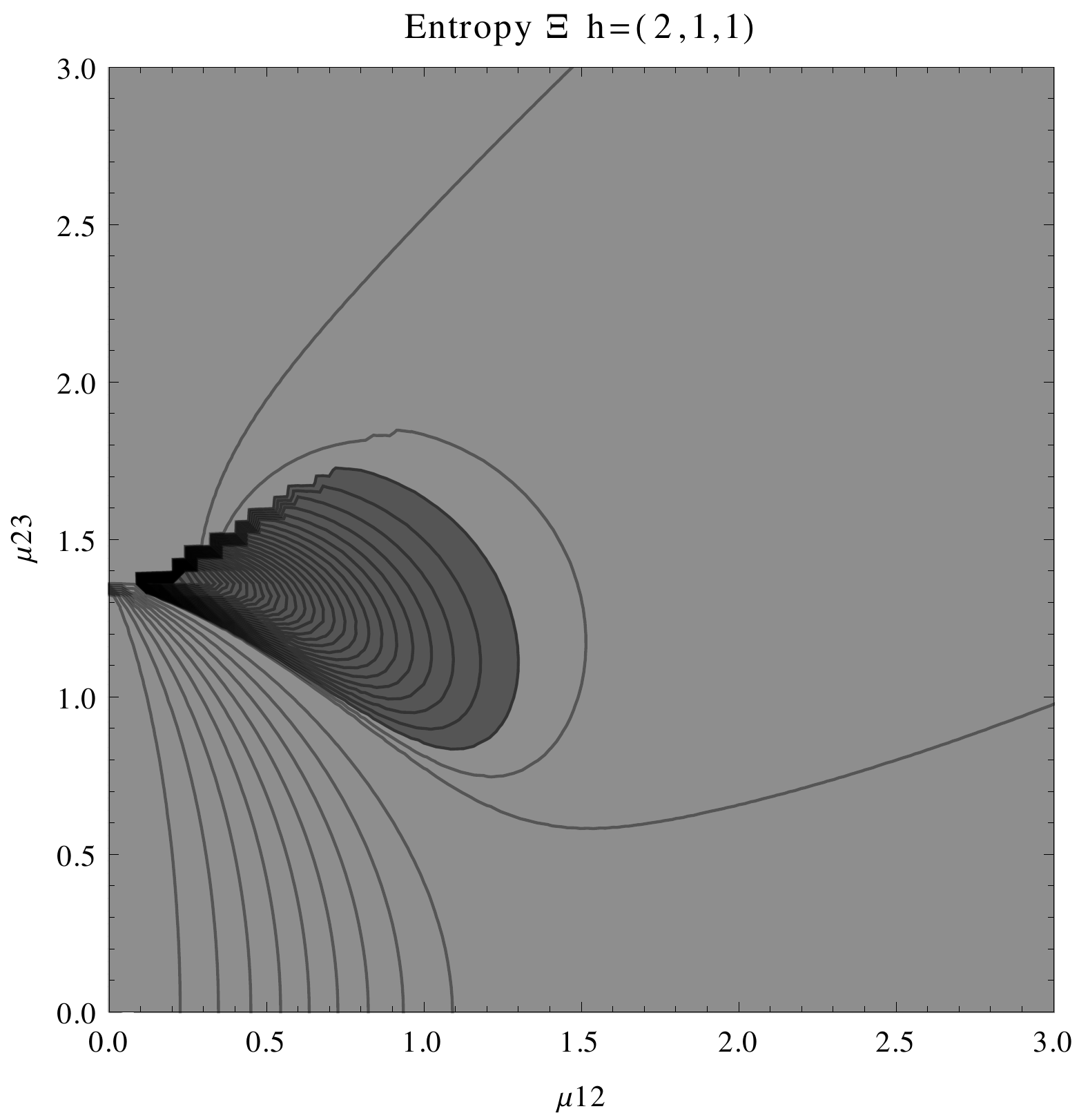}\quad{}\quad{}\includegraphics[scale=0.18]{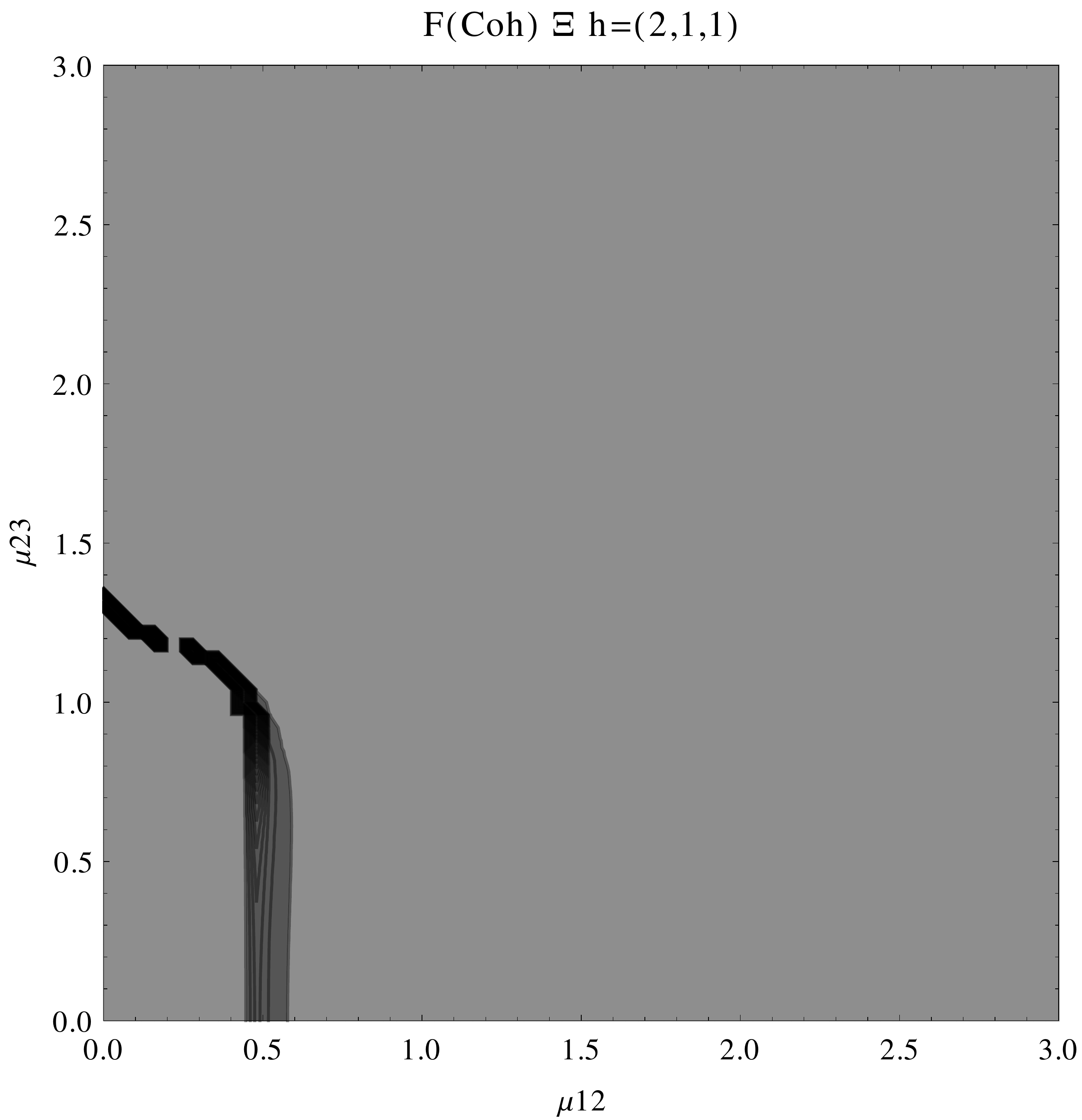}
\caption{(\textbf{Left}) 3D plot of the entropy of entanglement as a function of the coupling parameters $\mu_{12}$ and $\mu_{23}$, the maximum value of the entropy is $S_{\varepsilon}=1.32$ and the region where $S_{\varepsilon}>1.02$ is shown in dark grey. (\textbf{Center}) Contour plot of the entropy of entanglement as a function of the coupling parameters $\mu_{12}$ and $\mu_{23}$, the region where $S_{\varepsilon}>1.02$ is shown in dark grey. (\textbf{Right}) Fidelity between neighbouring coherent states as a function of the coupling parameters $\mu_{12}$ and $\mu_{23}$, dark grey region shows the fidelity’s minimum (i.e. the phase transition). All figures use $\omega_{1}=1.\bar{3}$, $\omega_{2}=1.\bar{6}$, $\Omega=0.5$ and correspond to the $\Xi$ configuration and the $h=(2,1,1)$ representation.}\label{fig:2}
\end{figure}

\begin{figure}
\centering
\includegraphics[scale=0.29]{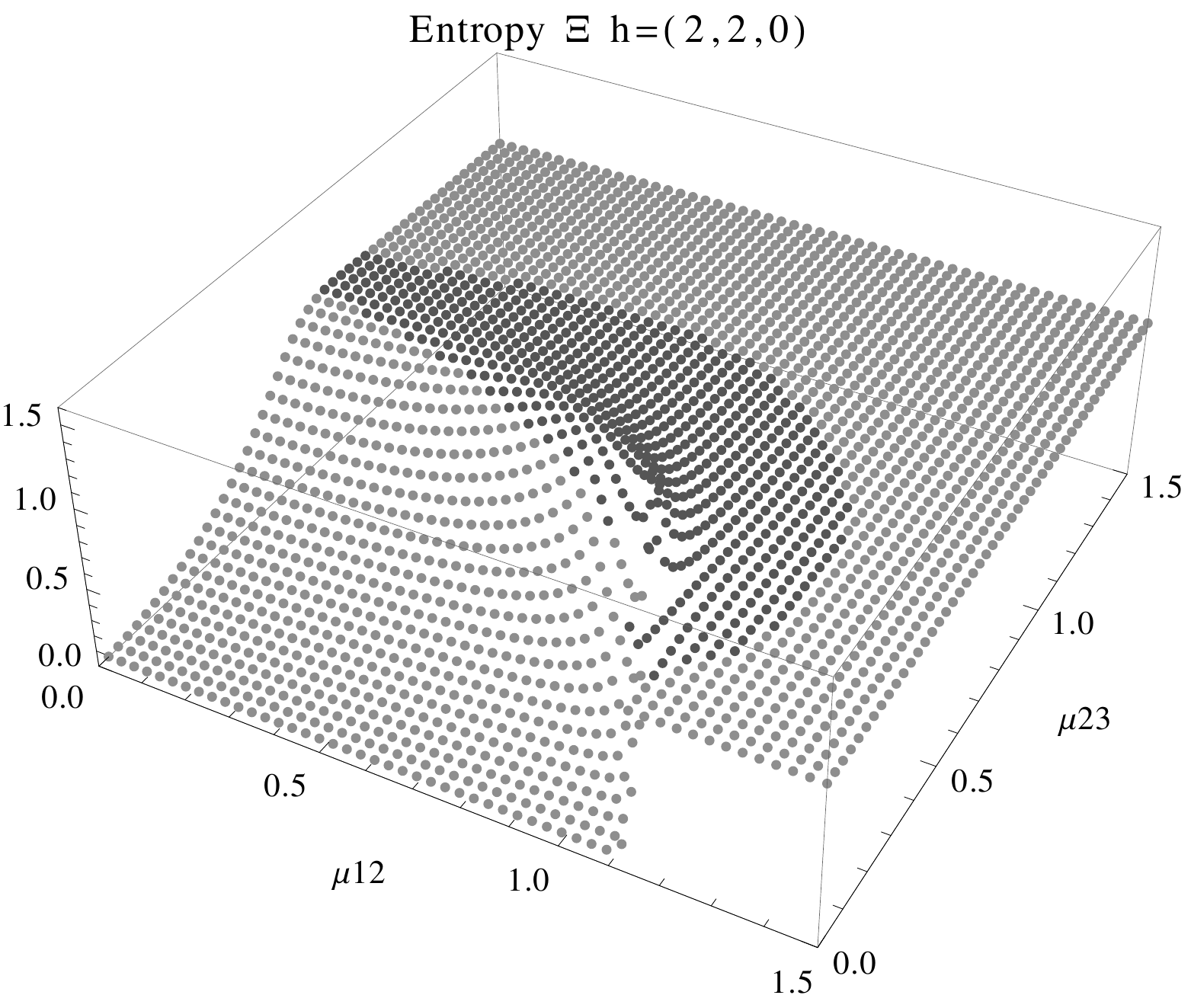}\quad{}\quad{}\includegraphics[scale=0.29]{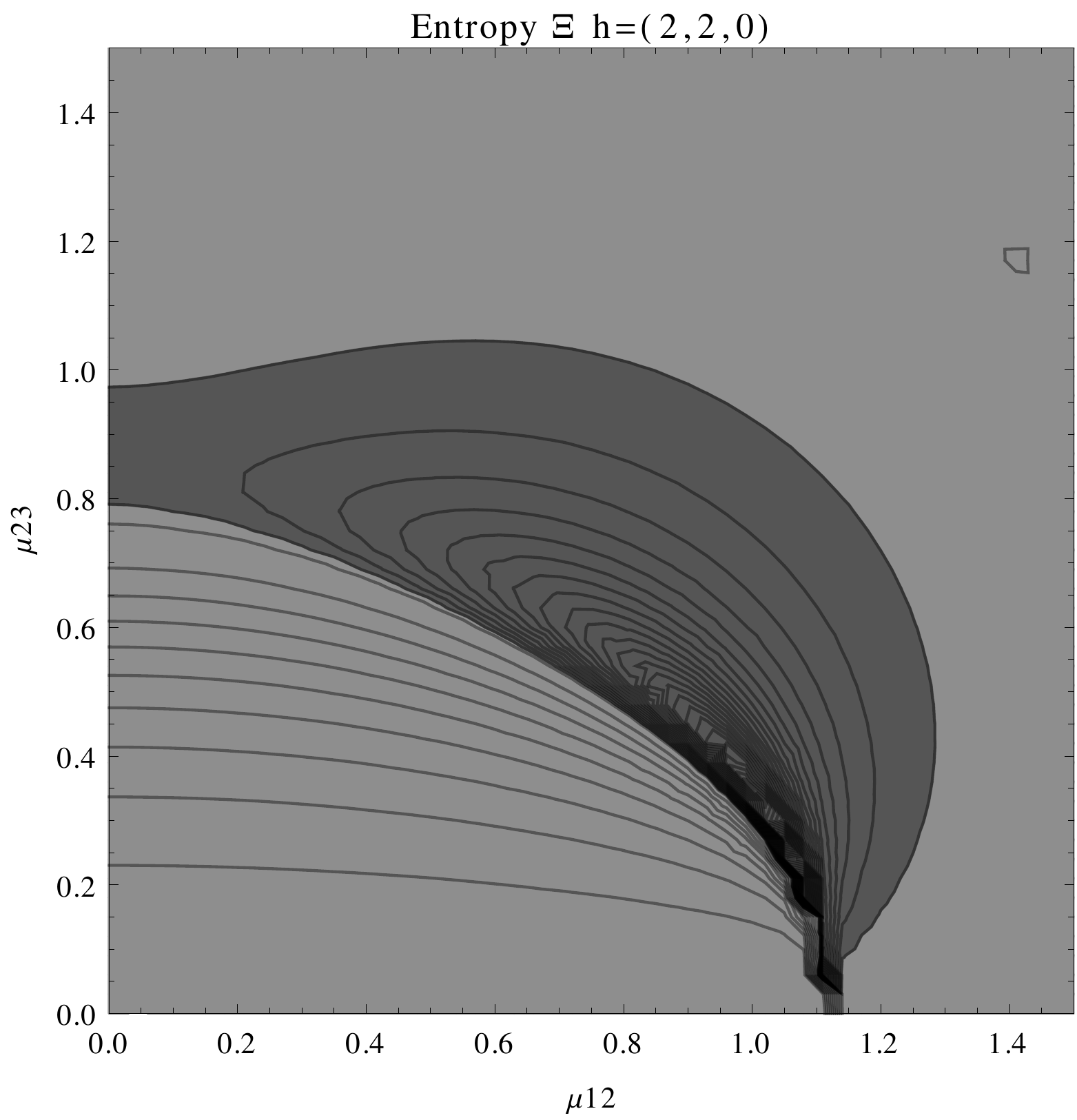}\quad{}\quad{}\includegraphics[scale=0.18]{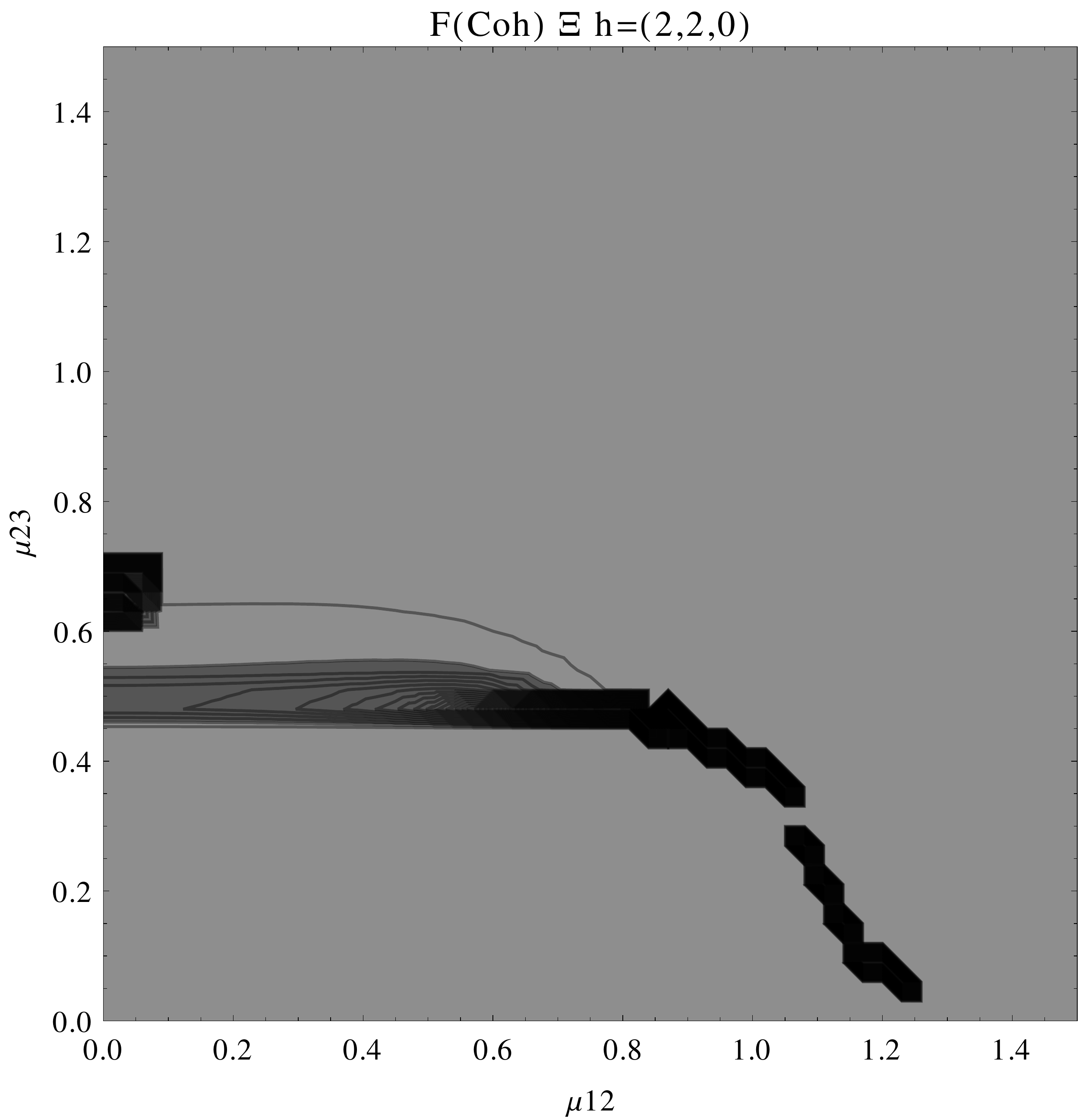}
\caption{(\textbf{Left}) 3D plot of the entropy of entanglement as a function of the coupling parameters $\mu_{12}$ and $\mu_{23}$, the maximum value of the entropy is $S_{\varepsilon}=1.58$ and the region where $S_{\varepsilon}>1.02$ is shown in dark grey. (\textbf{Center}) Contour plot of the entropy of entanglement as a function of the coupling parameters $\mu_{12}$ and $\mu_{23}$, the region where $S_{\varepsilon}>1.02$ is shown in dark grey. (\textbf{Right}) Fidelity between neighbouring coherent states as a function of the coupling parameters $\mu_{12}$ and $\mu_{23}$, dark grey region shows the fidelity’s minimum (i.e. the phase transition). All figures use $\omega_{1}=1.\bar{3}$, $\omega_{2}=1.\bar{6}$, $\Omega=0.5$ and correspond to the $\Xi$ configuration and the $h=(2,2,0)$ representation.}\label{fig:3}
\end{figure}

\begin{figure}
\centering
\includegraphics[scale=0.29]{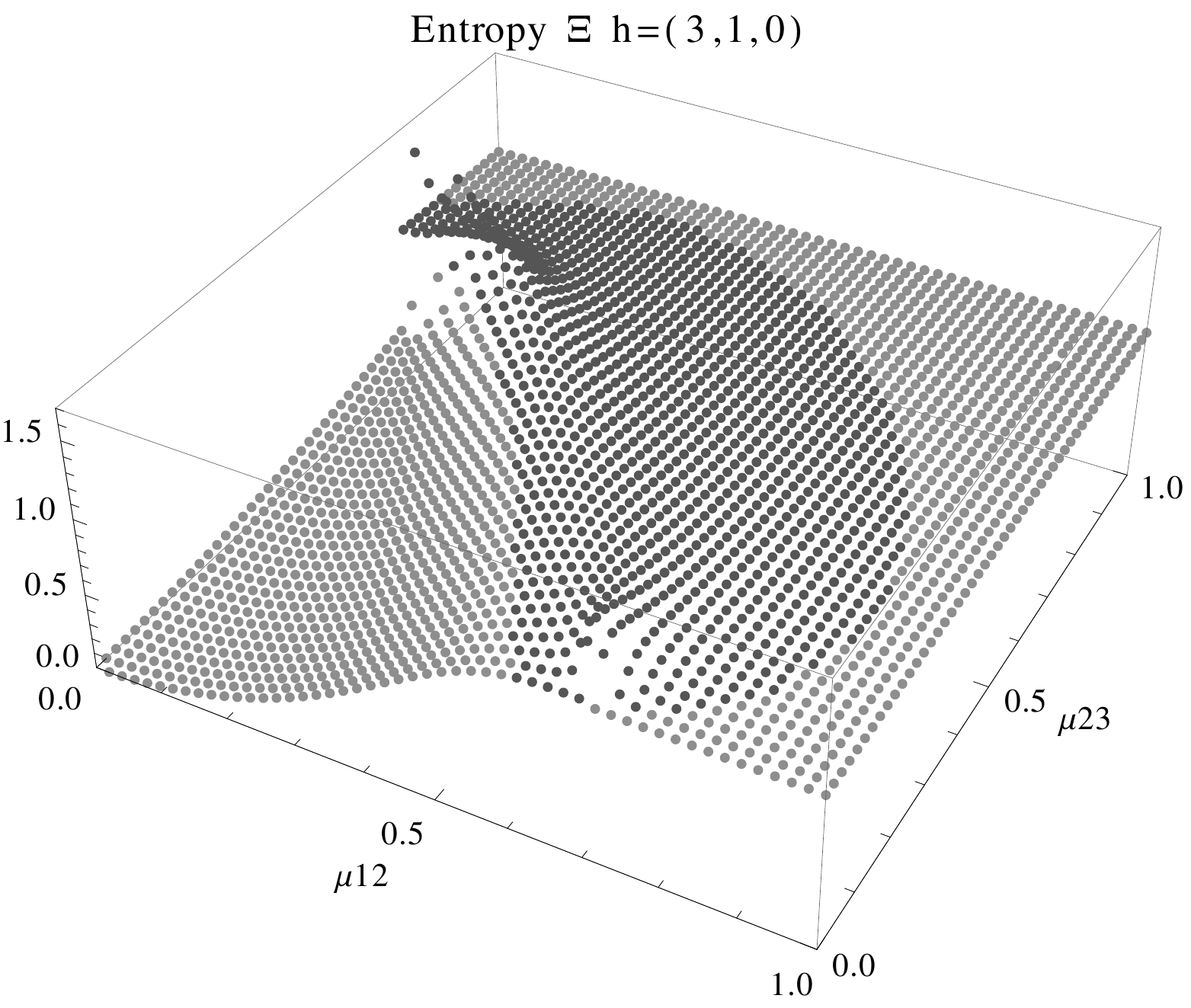}\quad{}\quad{}\includegraphics[scale=0.29]{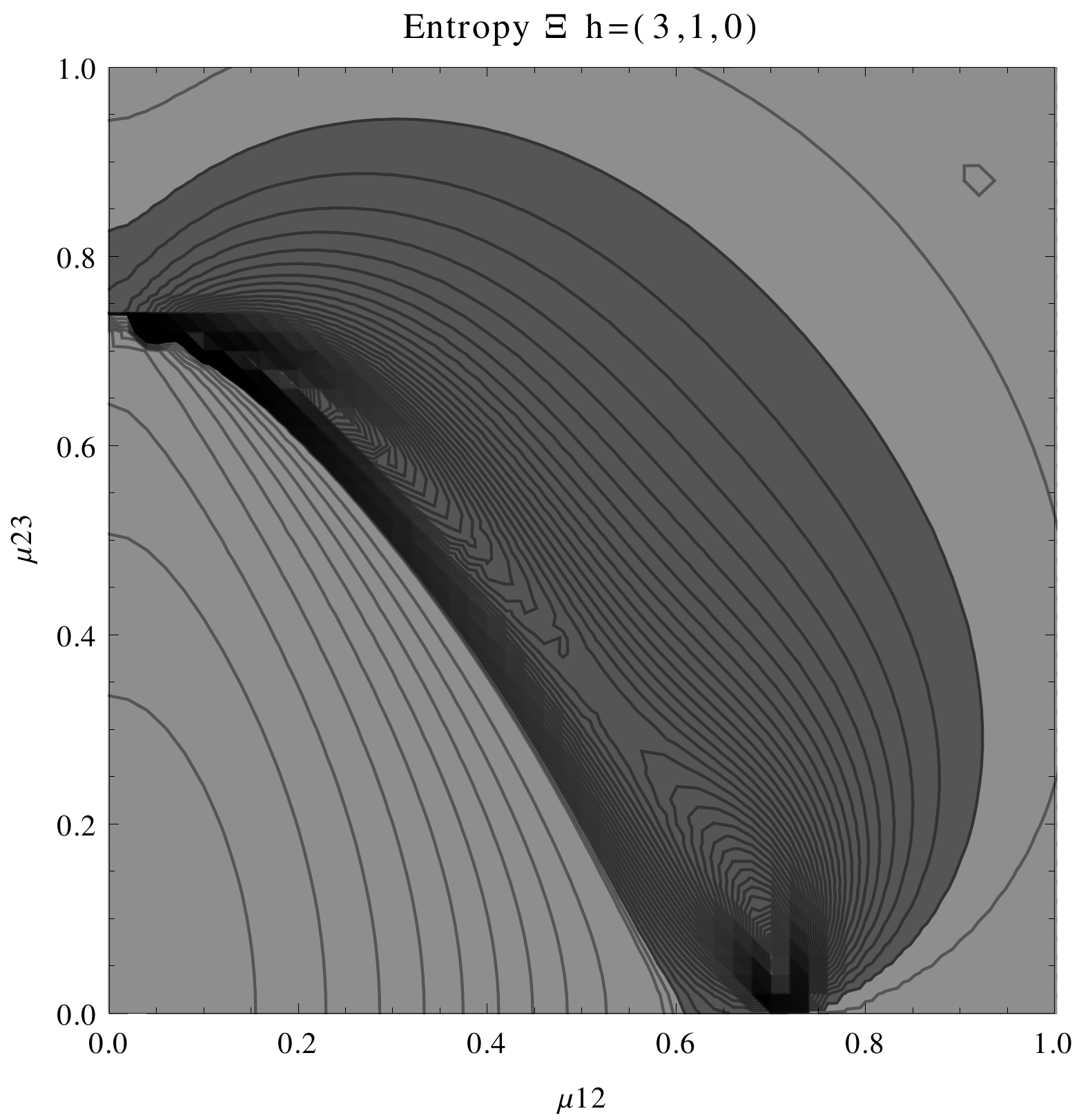}\quad{}\quad{}\includegraphics[scale=0.18]{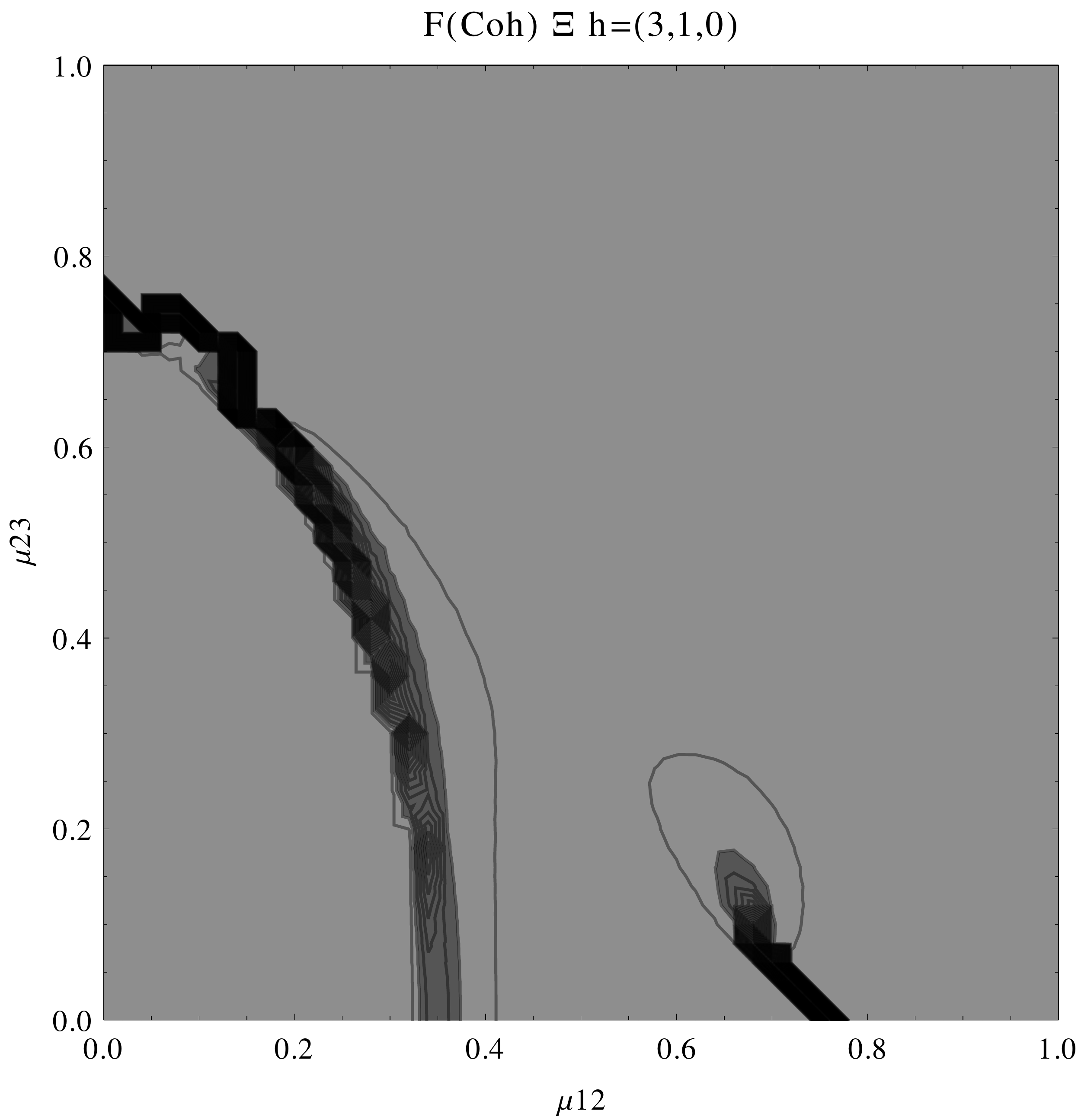}
\caption{(\textbf{Left}) 3D plot of the entropy of entanglement as a function of the coupling parameters $\mu_{12}$ and $\mu_{23}$, the maximum value of the entropy is $S_{\varepsilon}=1.65$ and the region where $S_{\varepsilon}>1.02$ is shown in dark grey. (\textbf{Center}) Contour plot of the entropy of entanglement as a function of the coupling parameters $\mu_{12}$ and $\mu_{23}$, the region where $S_{\varepsilon}>1.02$ is shown in dark grey. (\textbf{Right}) Fidelity between neighbouring coherent states as a function of the coupling parameters $\mu_{12}$ and $\mu_{23}$, dark grey region shows the fidelity’s minimum (i.e. the phase transition). All figures use $\omega_{1}=1.\bar{3}$, $\omega_{2}=1.\bar{6}$, $\Omega=0.5$ and correspond to the $\Xi$ configuration and the $h=(3,1,0)$ representation.}\label{fig:4}
\end{figure}

\begin{figure}
\centering
\includegraphics[scale=0.29]{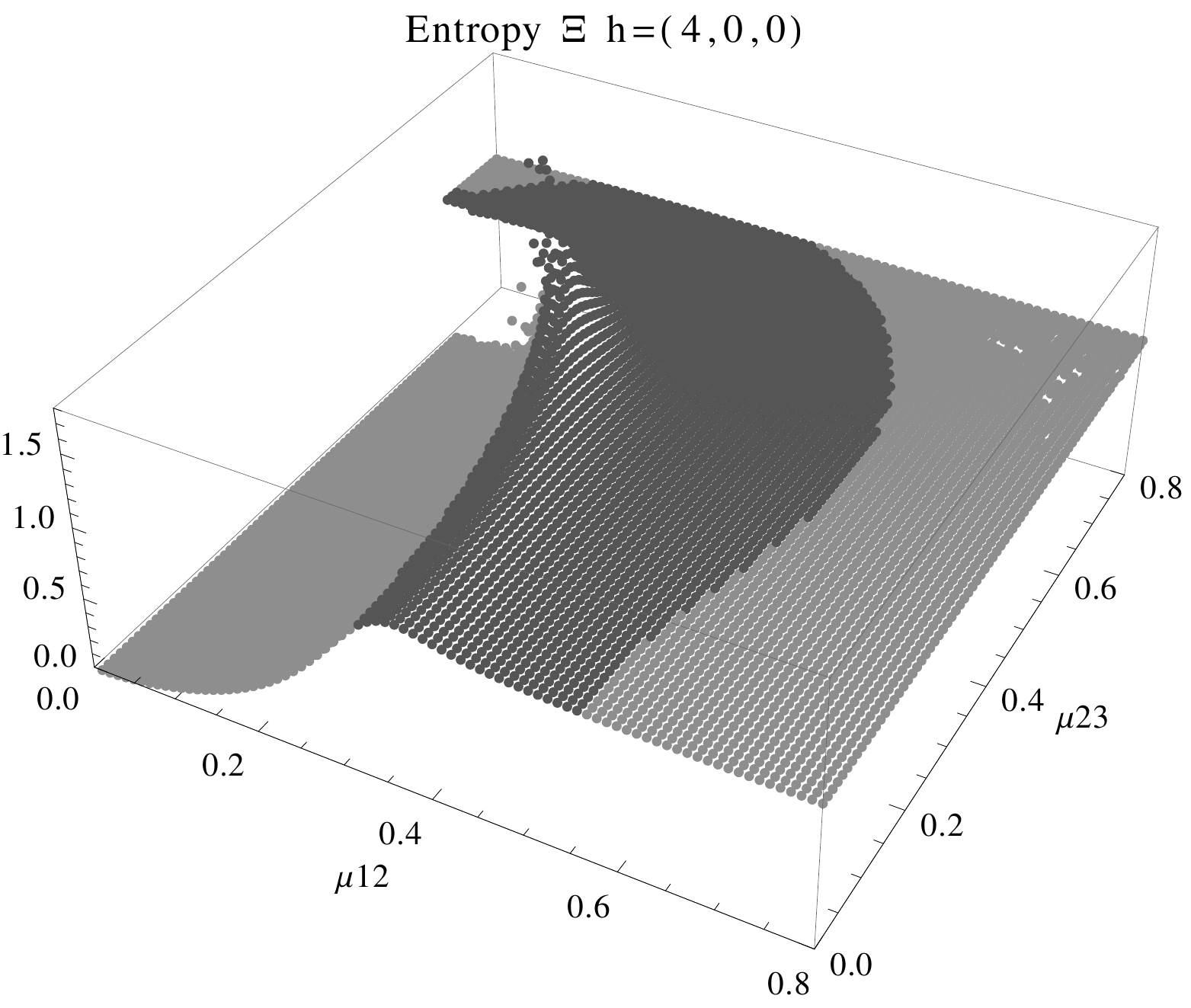}\quad{}\quad{}\includegraphics[scale=0.29]{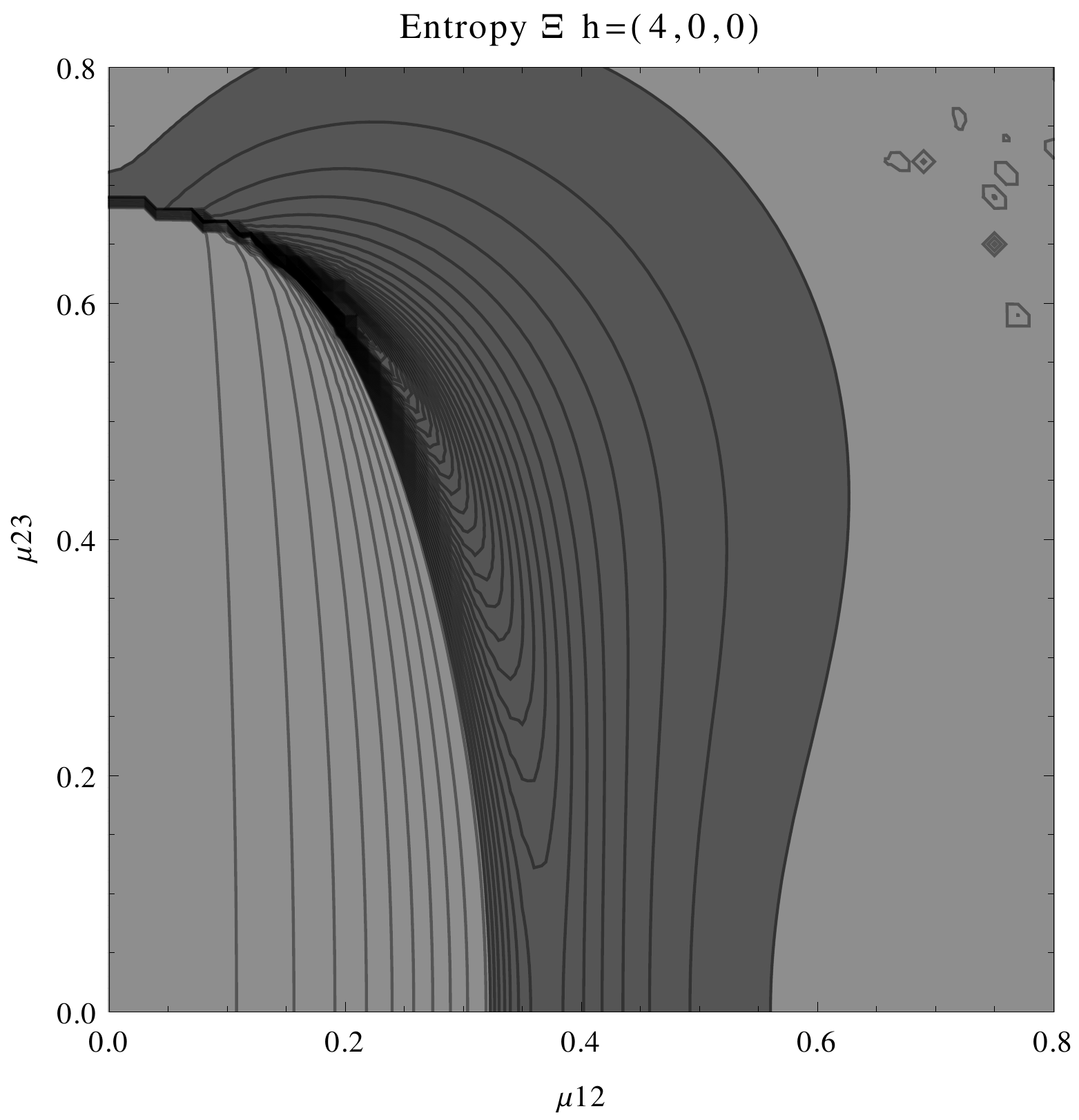}\quad{}\quad{}\includegraphics[scale=0.18]{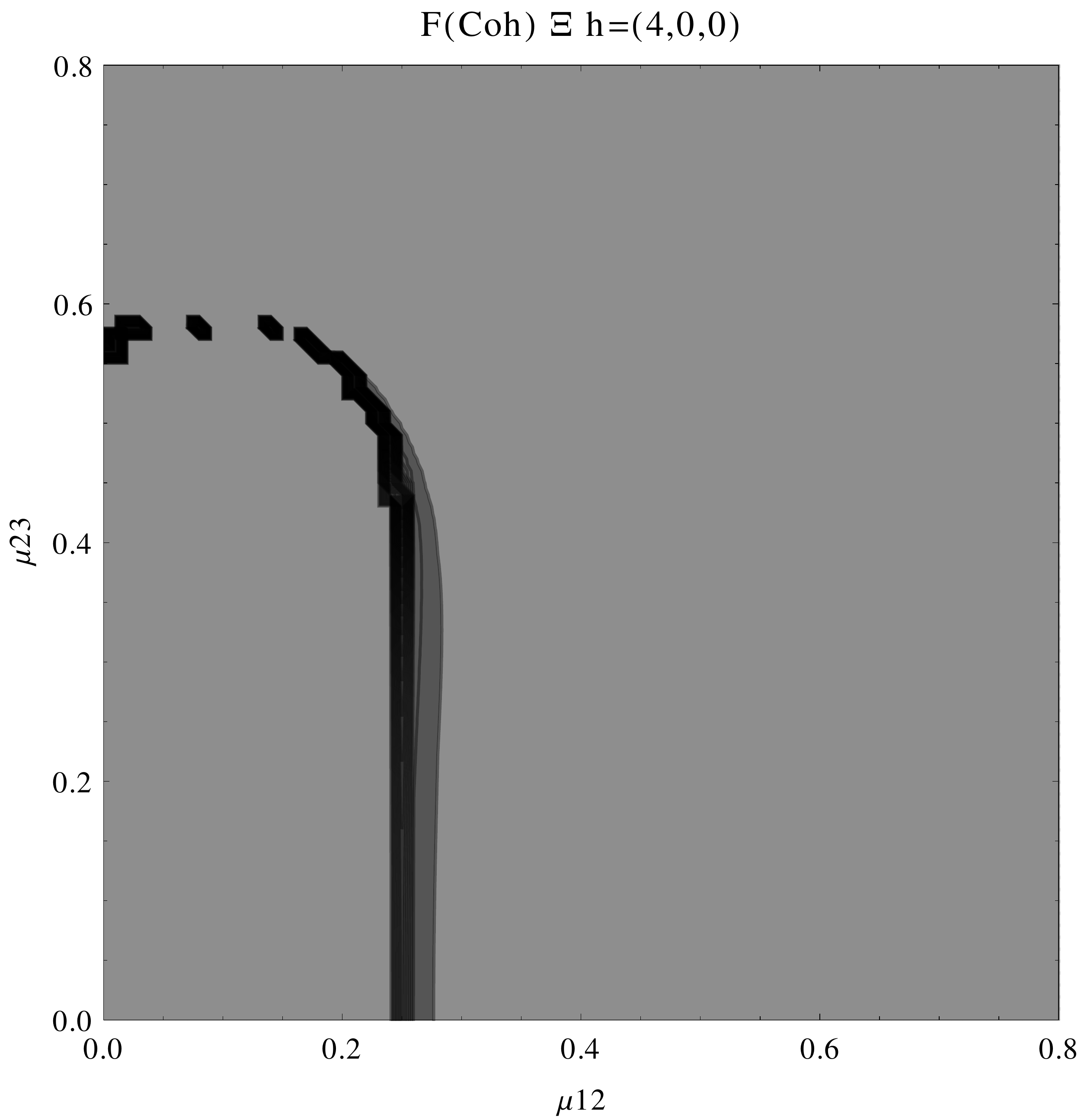}
\caption{(\textbf{Left}) 3D plot of the entropy of entanglement as a function of the coupling parameters $\mu_{12}$ and $\mu_{23}$, the maximum value of the entropy is $S_{\varepsilon}=1.78$ and the region where $S_{\varepsilon}>1.02$ is shown in dark grey. (\textbf{Center}) Contour plot of the entropy of entanglement as a function of the coupling parameters $\mu_{12}$ and $\mu_{23}$, the region where $S_{\varepsilon}>1.02$ is shown in dark grey. (\textbf{Right}) Fidelity between neighbouring coherent states as a function of the coupling parameters $\mu_{12}$ and $\mu_{23}$, dark grey region shows the fidelity’s minimum (i.e. the phase transition). All figures use $\omega_{1}=1.\bar{3}$, $\omega_{2}=1.\bar{6}$, $\Omega=0.5$ and correspond to the $\Xi$ configuration and the $h=(4,0,0)$ representation.}\label{fig:5}
\end{figure}

Results for the atoms being in the $\Xi$ configuration are presented in figures \ref{fig:2} to \ref{fig:5} for all four possible cooperation numbers. The first two graphics (from left to right) show the entropy of entanglement. In them, the region where the entropy reaches its highest values ($S_{\varepsilon}>1.02$) is shown in dark grey. It is worth noting that this region gets larger as the cooperation number increases.

The third graphic shows a contour plot of the fidelity between neighbouring coherent states. In this, the region where the fidelity drops ($F<0.97$ is emphasised although fidelity drops to values near zero) is shown in dark grey. Irregularities appear due to numerical errors in the energy surface's minimisation process near the transition.

\begin{figure}
\centering
\includegraphics[scale=0.29]{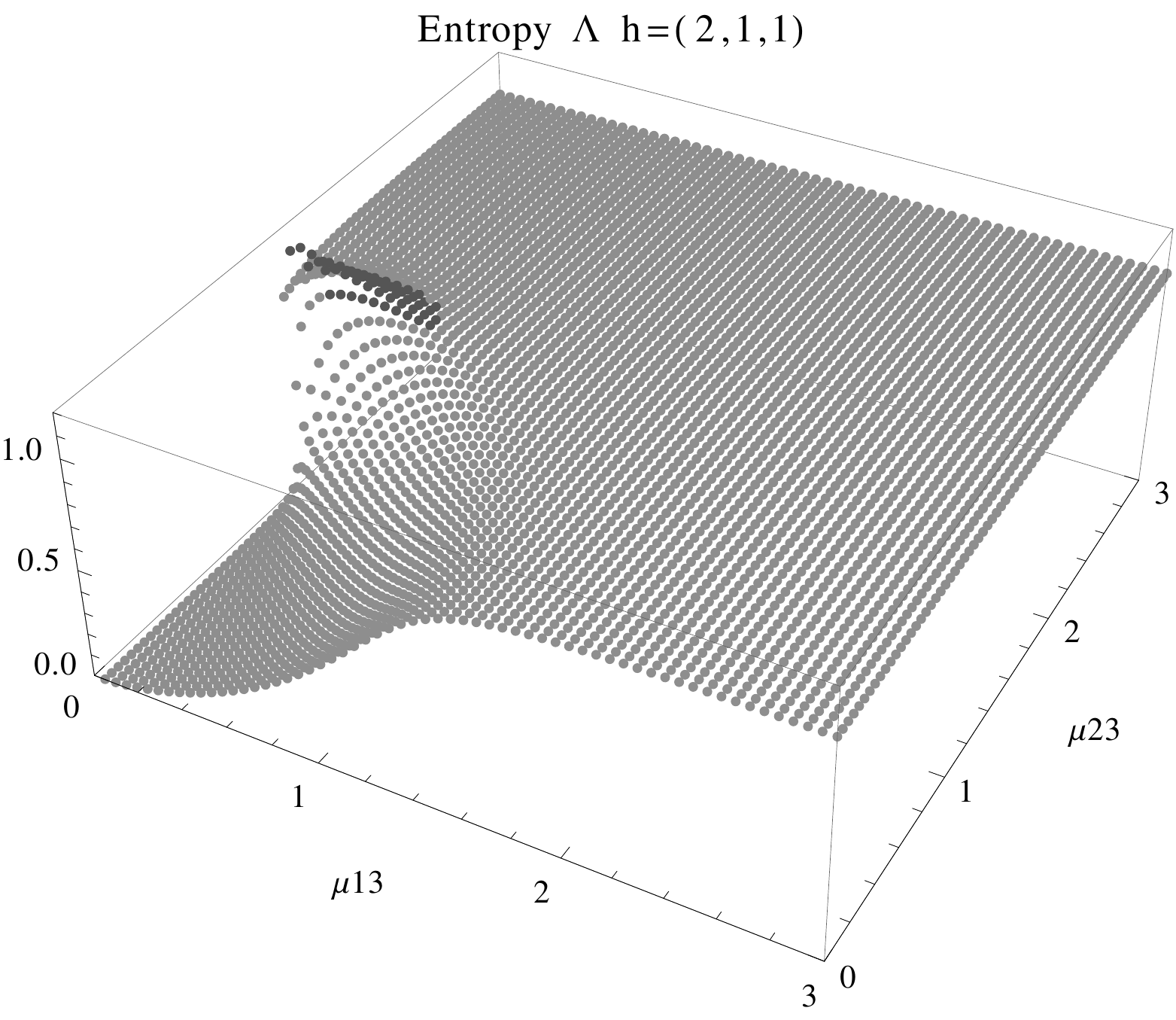}\quad{}\quad{}\includegraphics[scale=0.29]{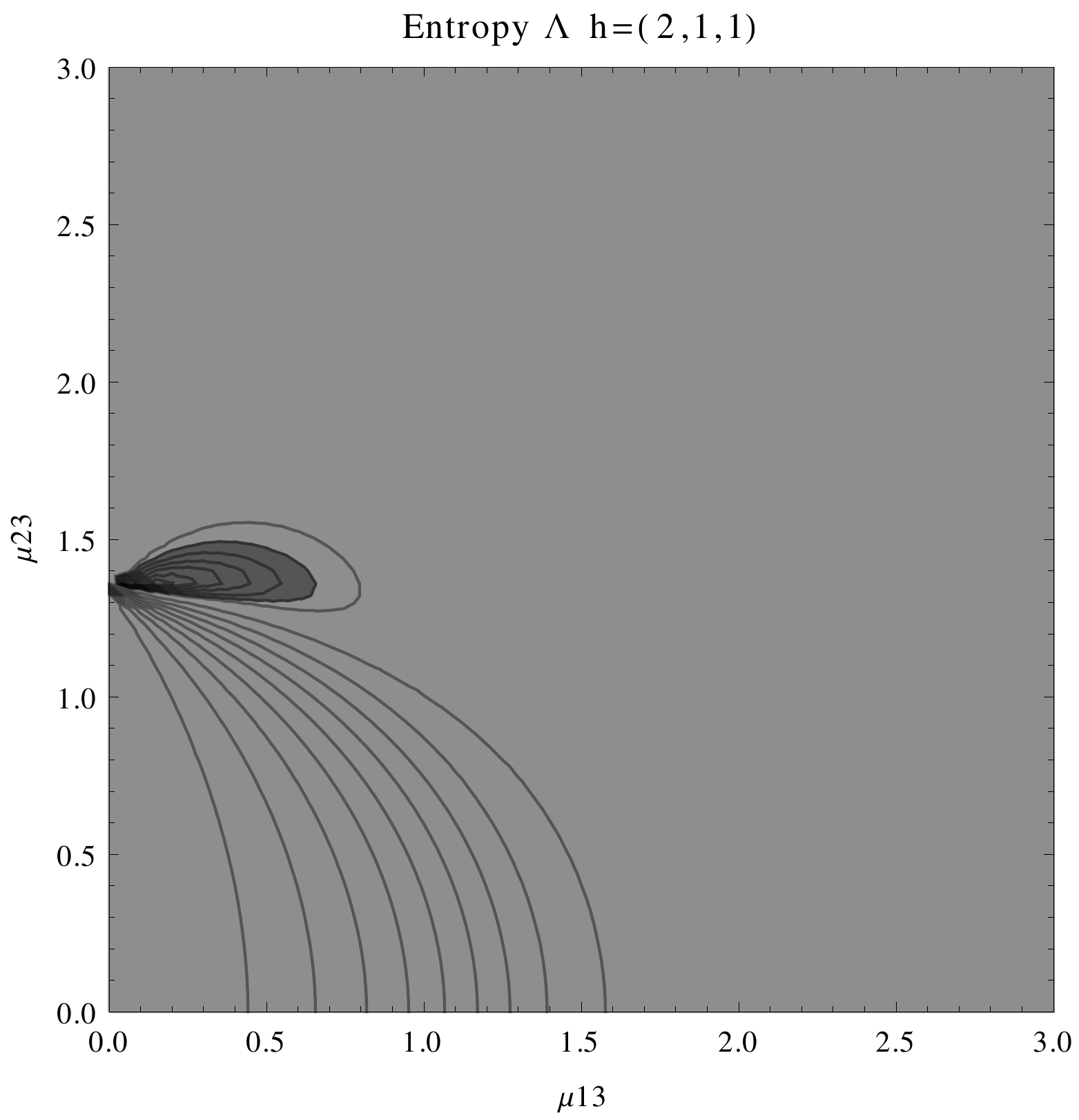}\quad{}\quad{}\includegraphics[scale=0.18]{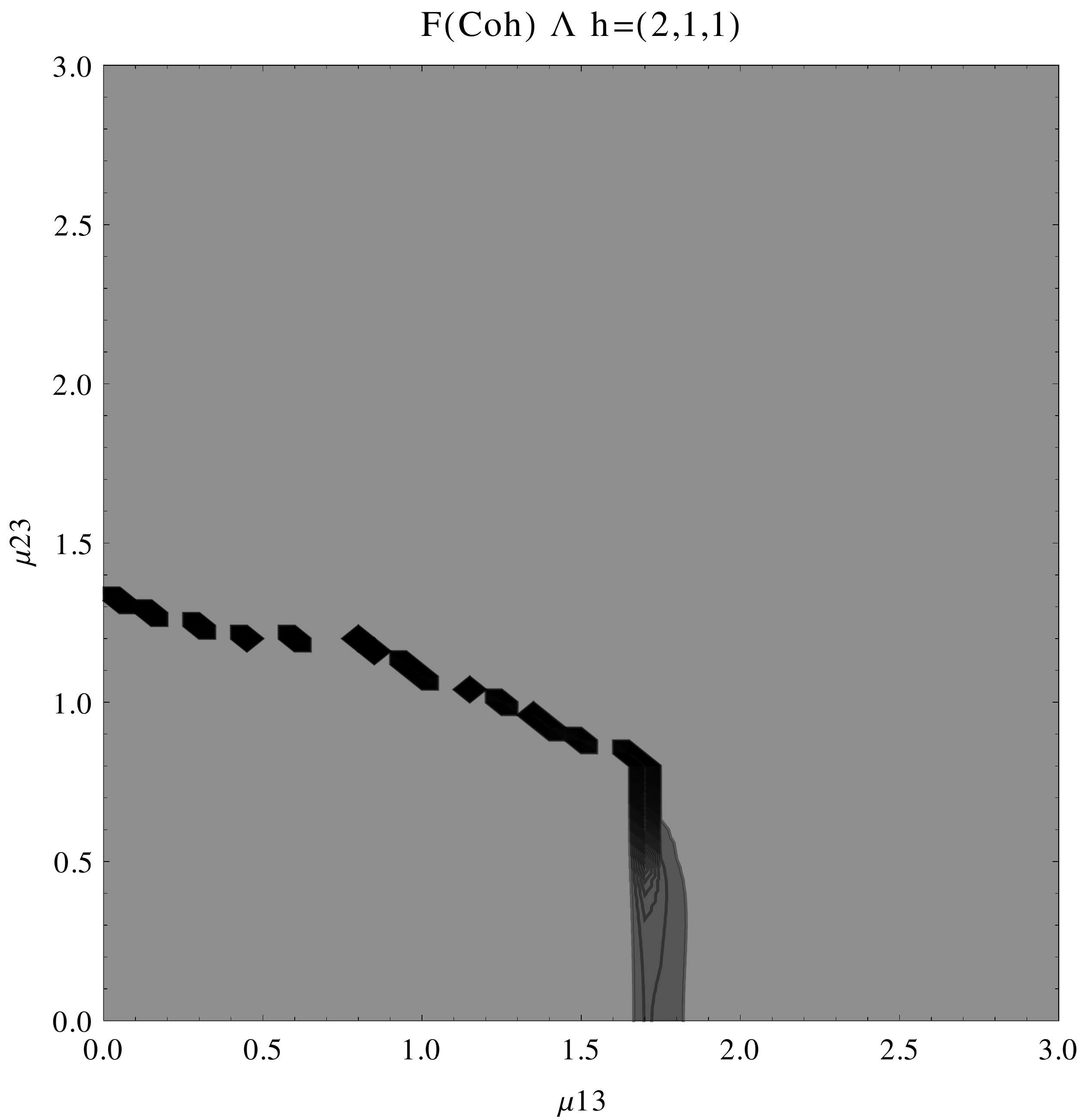}
\caption{(\textbf{Left}) 3D plot of the entropy of entanglement as a function of the coupling parameters $\mu_{13}$ and $\mu_{23}$, the maximum value of the entropy is $S_{\varepsilon}=1.15$ and the region where $S_{\varepsilon}>1.01$ is shown in dark grey. (\textbf{Center}) Contour plot of the entropy of entanglement as a function of the coupling parameters $\mu_{13}$ and $\mu_{23}$, the region where $S_{\varepsilon}>1.01$ is shown in dark grey. (\textbf{Right}) Fidelity between neighbouring coherent states as a function of the coupling parameters $\mu_{13}$ and $\mu_{23}$, dark grey region shows the fidelity’s minimum (i.e. the phase transition). All figures use $\omega_{1}=1.\bar{3}$, $\omega_{2}=1.\bar{6}$, $\Omega=0.5$ and correspond to the $\Lambda$ configuration and the $h=(2,1,1)$ representation.}\label{fig:6}
\end{figure}

\begin{figure}
\centering
\includegraphics[scale=0.24]{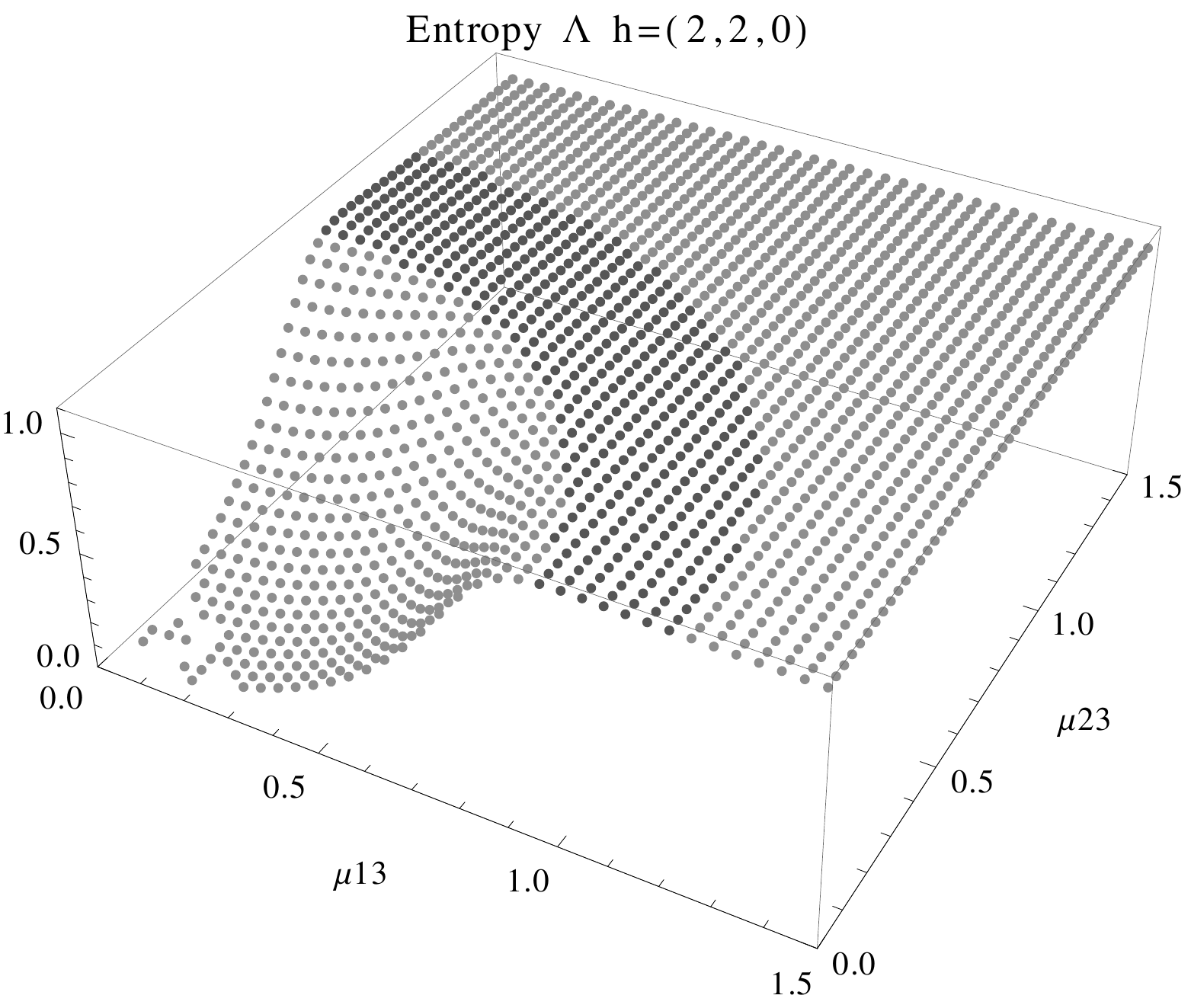}\quad{}\quad{}\includegraphics[scale=0.24]{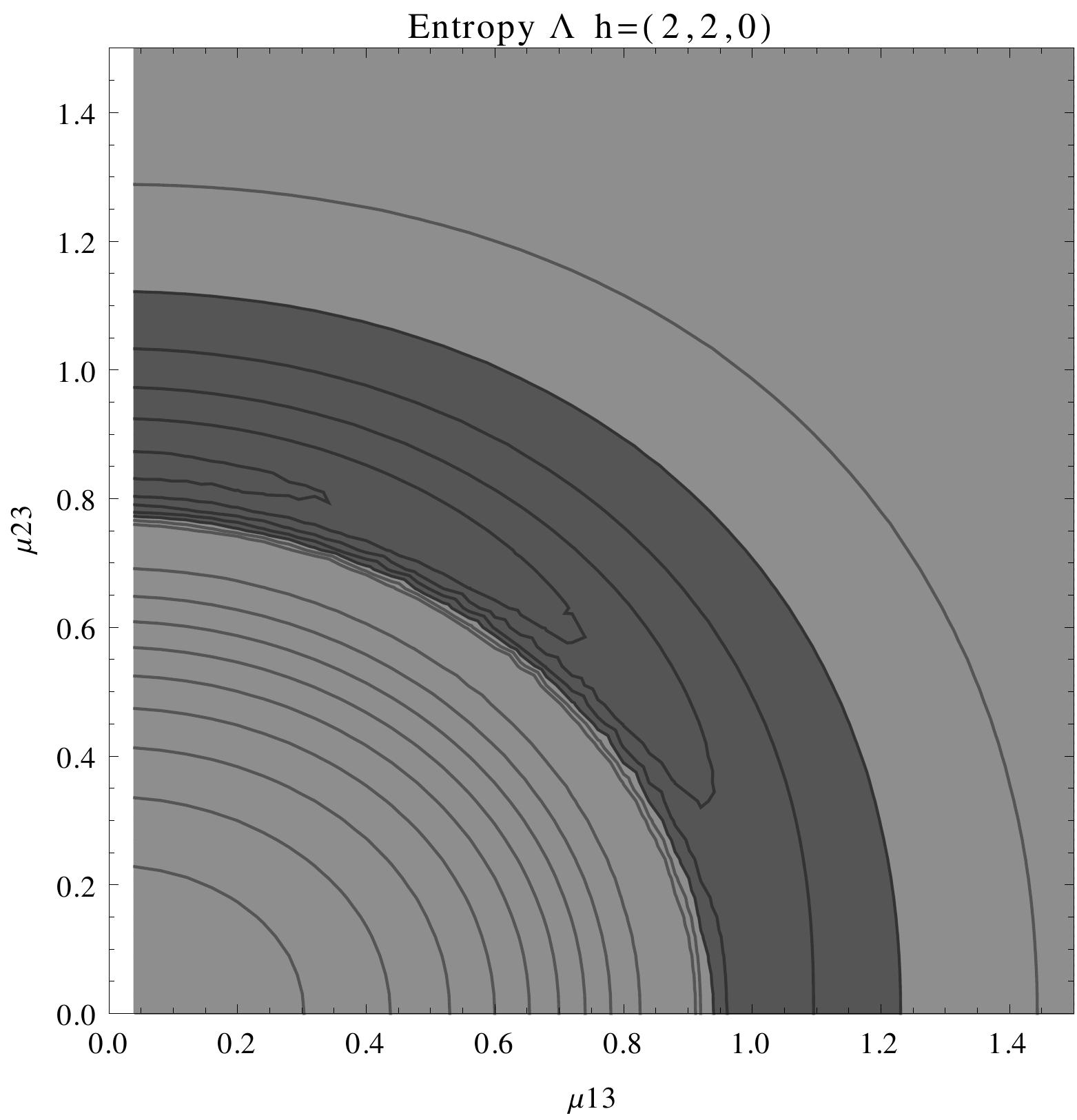}\quad{}\quad{}\includegraphics[scale=0.15]{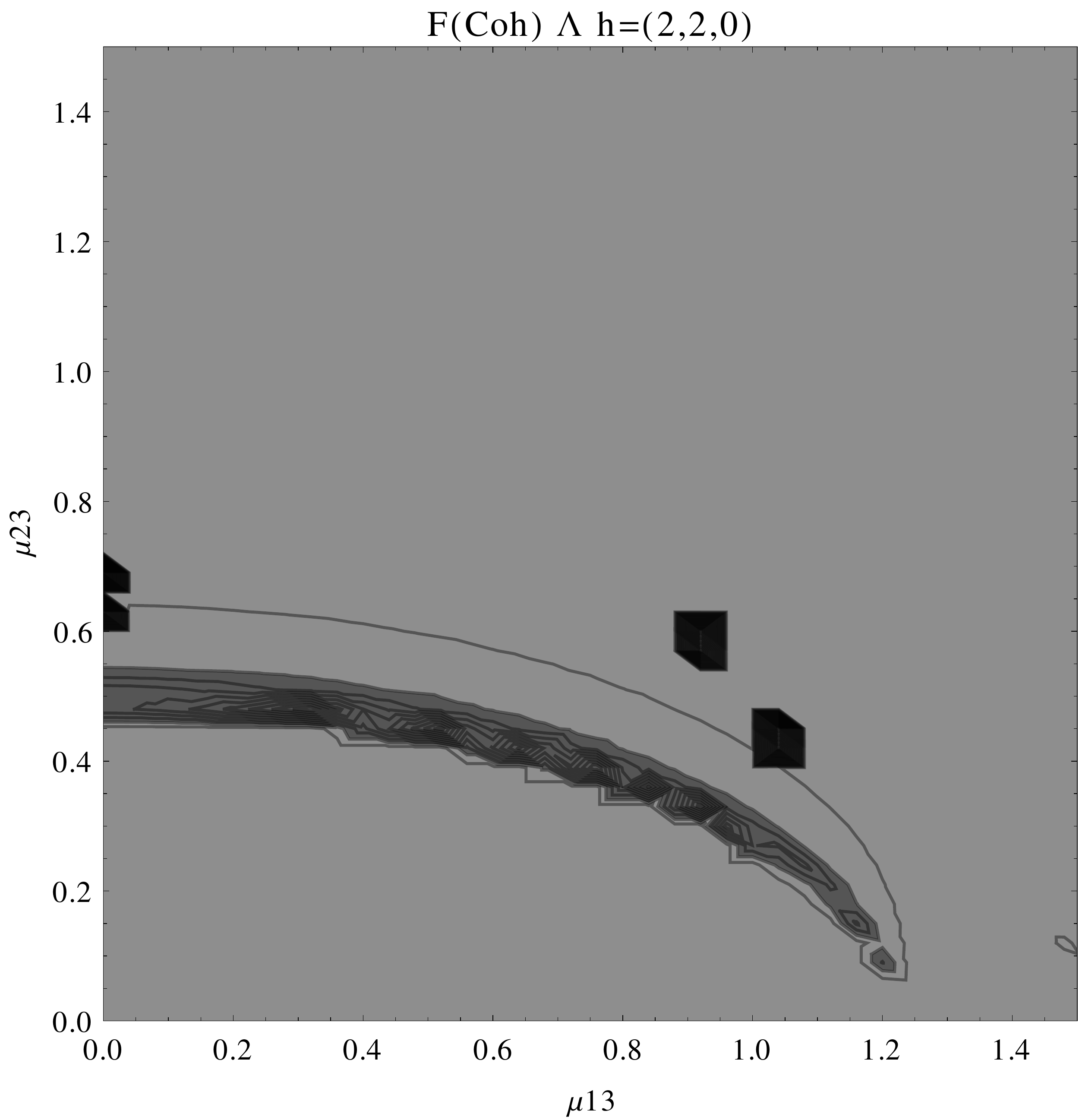}
\caption{(\textbf{Left}) 3D plot of the entropy of entanglement as a function of the coupling parameters $\mu_{13}$ and $\mu_{23}$, the maximum value of the entropy is $S_{\varepsilon}=1.03$ and the region where $S_{\varepsilon}>1.01$ is shown in dark grey. (\textbf{Center}) Contour plot of the entropy of entanglement as a function of the coupling parameters $\mu_{13}$ and $\mu_{23}$, the region where $S_{\varepsilon}>1.01$ is shown in dark grey. (\textbf{Right}) Fidelity between neighbouring coherent states as a function of the coupling parameters $\mu_{13}$ and $\mu_{23}$, dark grey region shows the fidelity’s minimum (i.e. the phase transition). All figures use $\omega_{1}=1.\bar{3}$, $\omega_{2}=1.\bar{6}$, $\Omega=0.5$ and correspond to the $\Lambda$ configuration and the $h=(2,2,0)$ representation.}\label{fig:7}
\end{figure}

\begin{figure}
\centering
\includegraphics[scale=0.24]{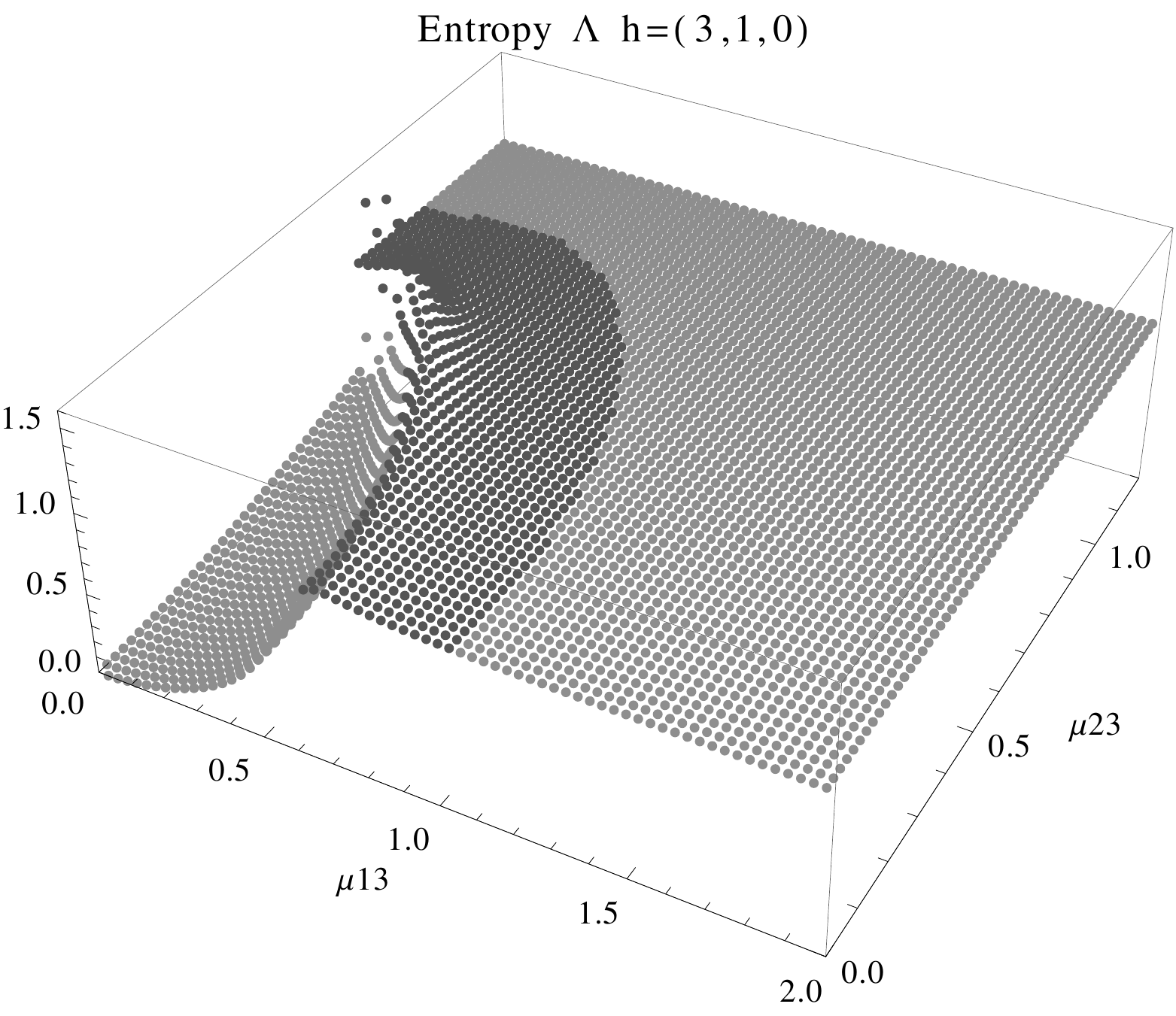}\quad{}\quad{}\includegraphics[scale=0.24]{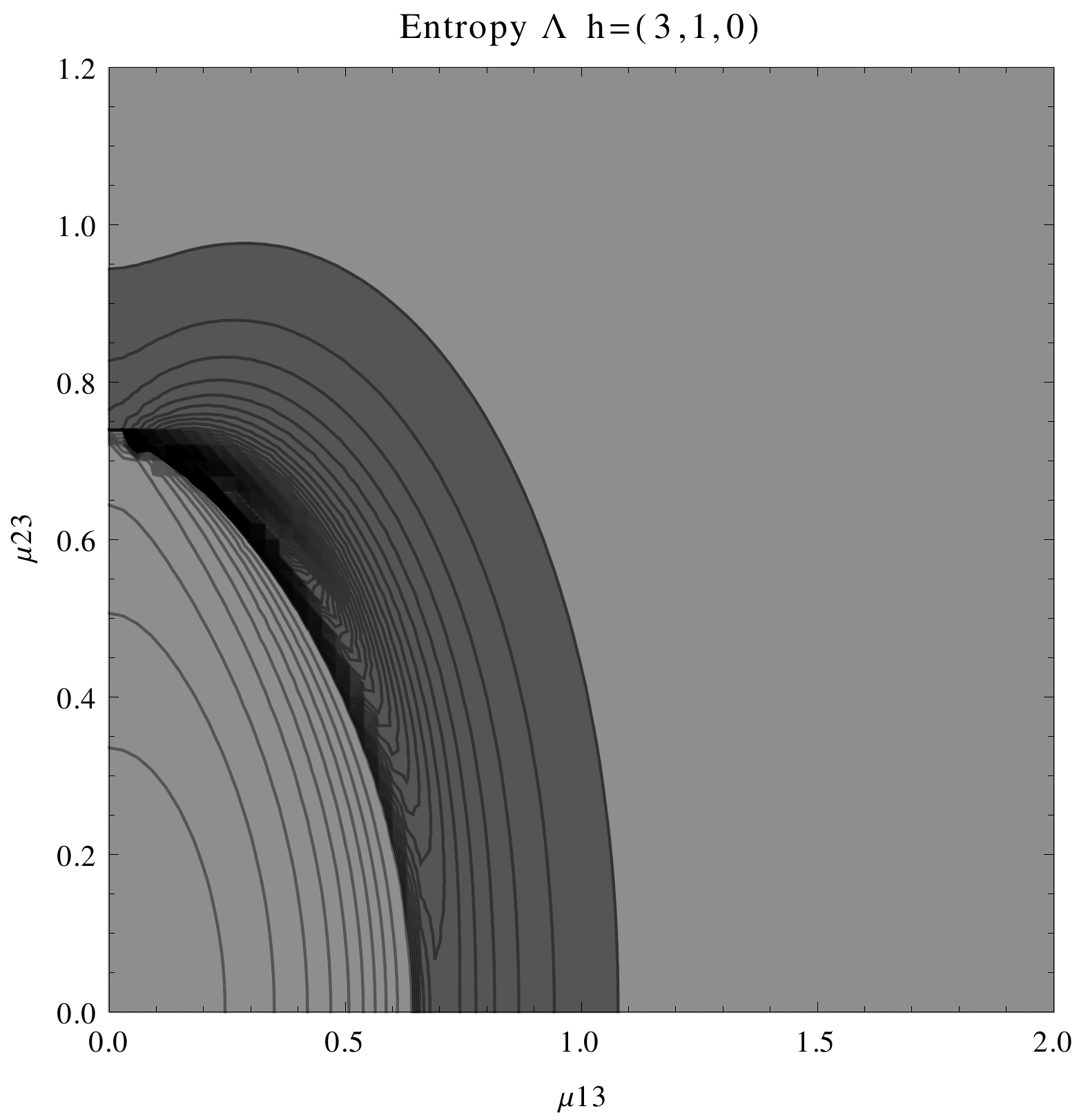}\quad{}\quad{}\includegraphics[scale=0.15]{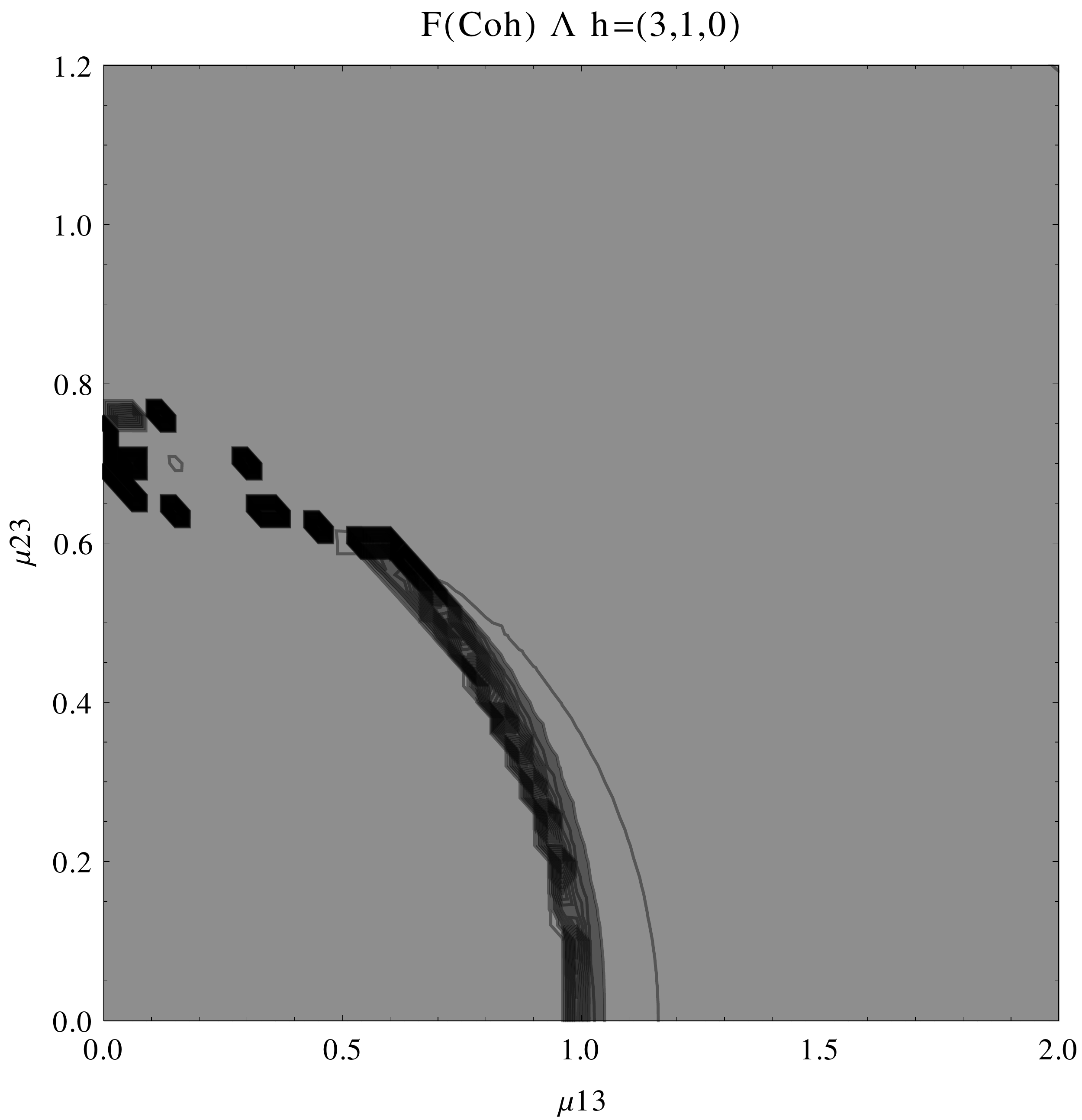}
\caption{(\textbf{Left}) 3D plot of the entropy of entanglement as a function of the coupling parameters $\mu_{13}$ and $\mu_{23}$, the maximum value of the entropy is $S_{\varepsilon}=1.59$ and the region where $S_{\varepsilon}>1.01$ is shown in dark grey. (\textbf{Center}) Contour plot of the entropy of entanglement as a function of the coupling parameters $\mu_{13}$ and $\mu_{23}$, the region where $S_{\varepsilon}>1.01$ is shown in dark grey. (\textbf{Right}) Fidelity between neighbouring coherent states as a function of the coupling parameters $\mu_{13}$ and $\mu_{23}$, dark grey region shows the fidelity’s minimum (i.e. the phase transition). All figures use $\omega_{1}=1.\bar{3}$, $\omega_{2}=1.\bar{6}$, $\Omega=0.5$ and correspond to the $\Lambda$ configuration and the $h=(3,1,0)$ representation.}\label{fig:8}
\end{figure}

\begin{figure}
\centering
\includegraphics[scale=0.24]{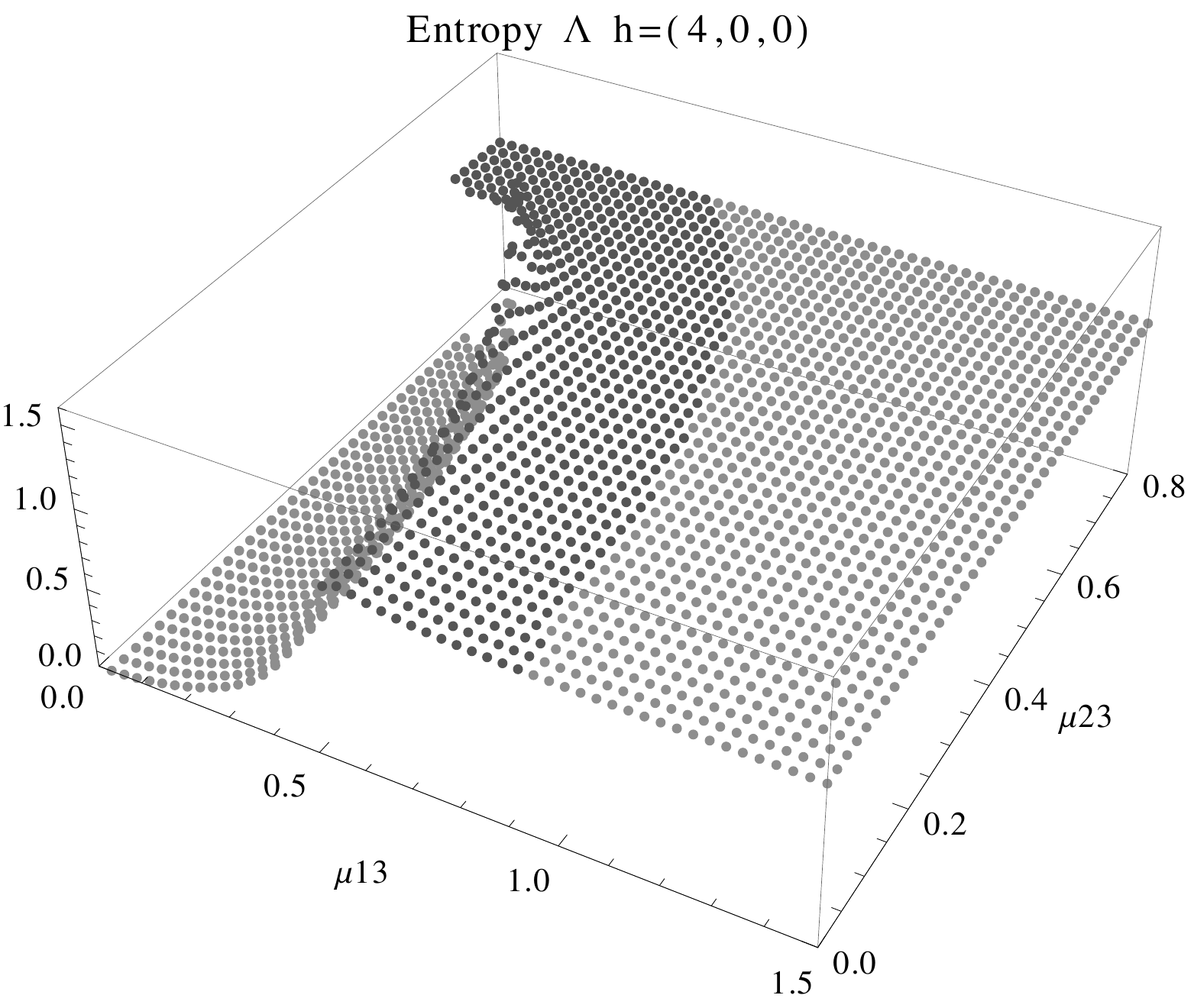}\quad{}\quad{}\includegraphics[scale=0.24]{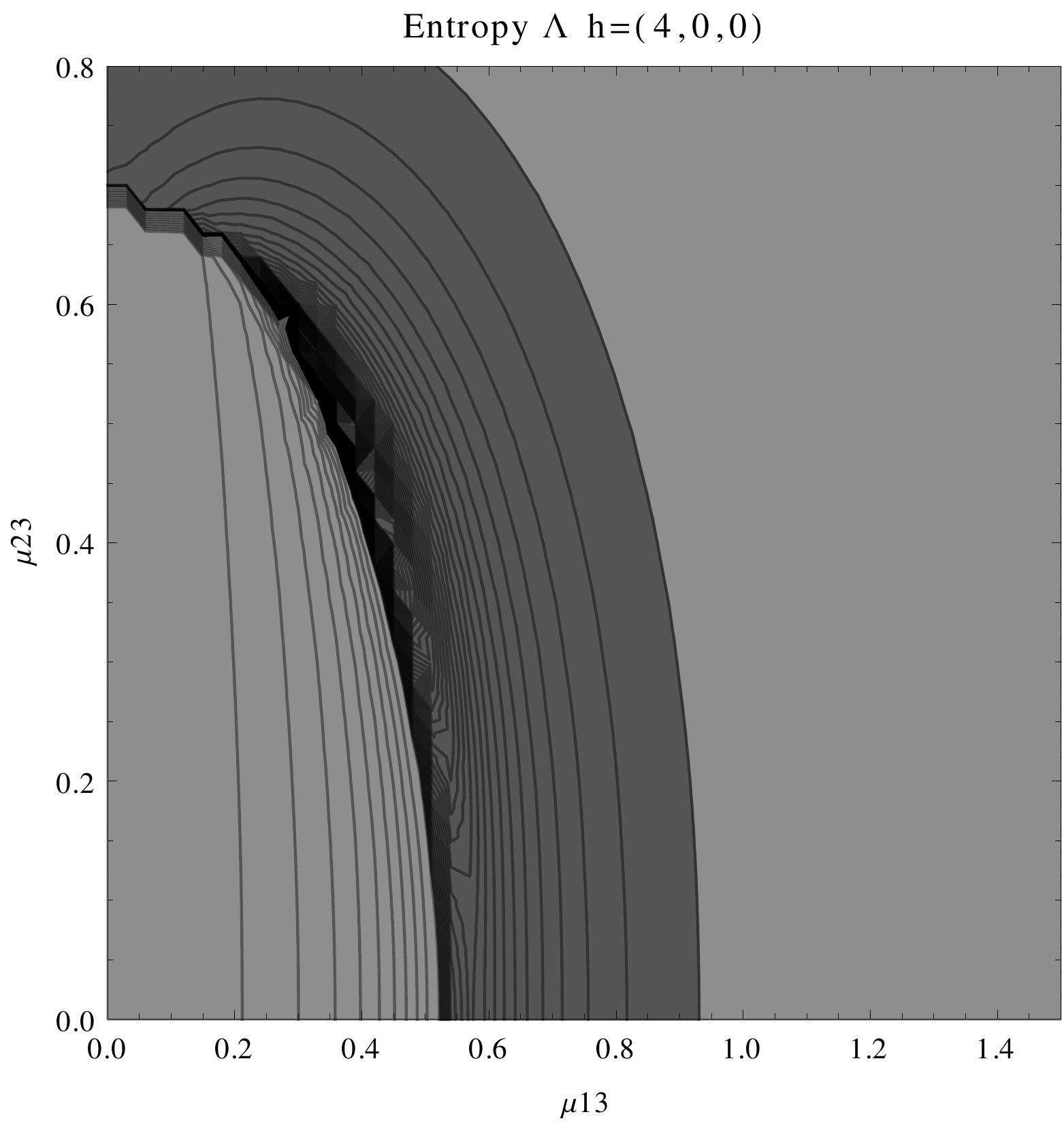}\quad{}\quad{}\includegraphics[scale=0.15]{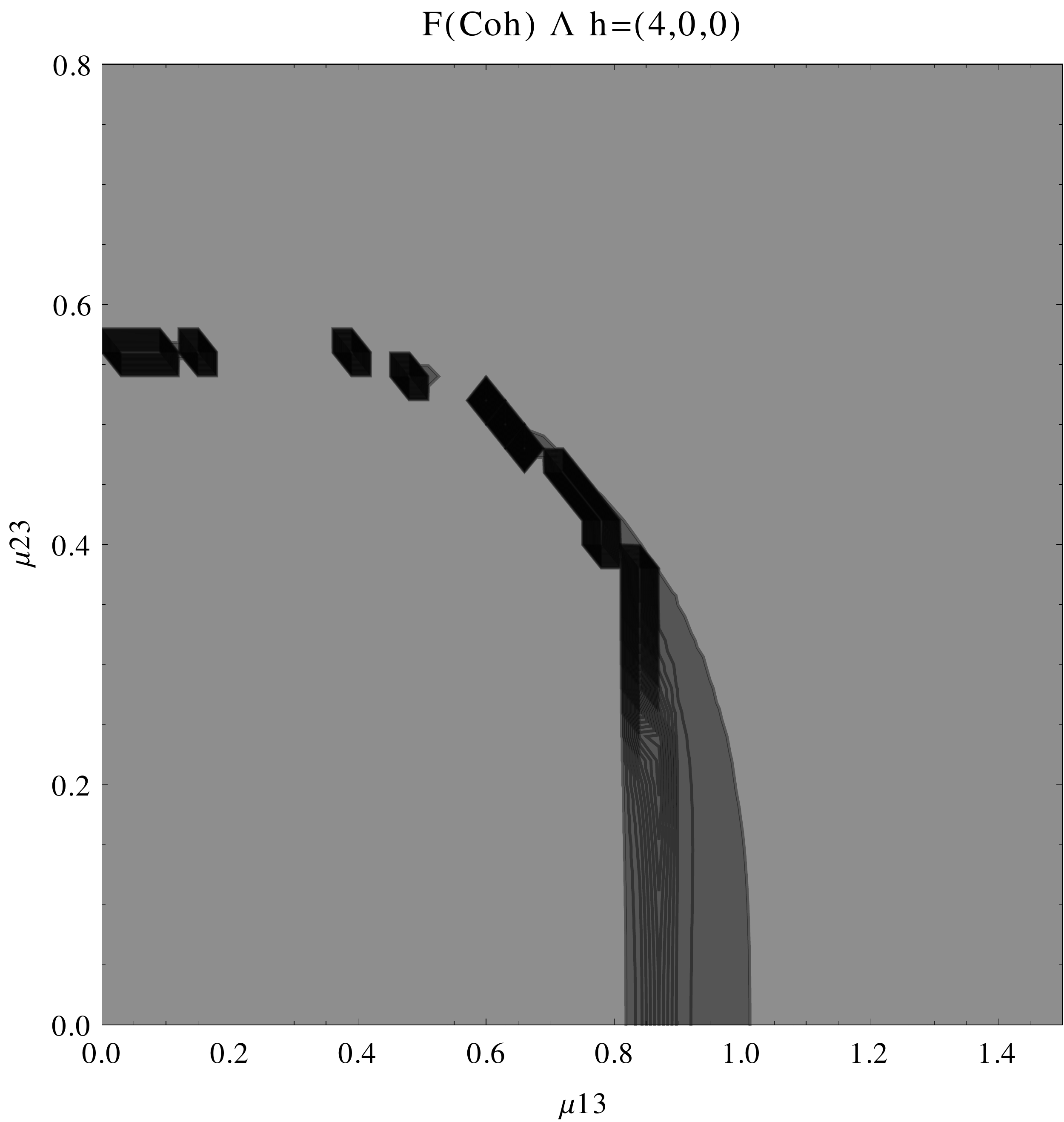}
\caption{(\textbf{Left}) 3D plot of the entropy of entanglement as a function of the coupling parameters $\mu_{13}$ and $\mu_{23}$, the maximum value of the entropy is $S_{\varepsilon}=1.55$ and the region where $S_{\varepsilon}>1.01$ is shown in dark grey. (\textbf{Center}) Contour plot of the entropy of entanglement as a function of the coupling parameters $\mu_{13}$ and $\mu_{23}$, the region where $S_{\varepsilon}>1.01$ is shown in dark grey. (\textbf{Right}) Fidelity between neighbouring coherent states as a function of the coupling parameters $\mu_{13}$ and $\mu_{23}$, dark grey region shows the fidelity’s minimum (i.e. the phase transition). All figures use $\omega_{1}=1.\bar{3}$, $\omega_{2}=1.\bar{6}$, $\Omega=0.5$ and correspond to the $\Lambda$ configuration and the $h=(4,0,0)$ representation.}\label{fig:9}
\end{figure}

The results for atoms in the $\Lambda$ configuration are presented in figures \ref{fig:6} to \ref{fig:9} for all four possible cooperation numbers. The first two graphics (from left to right) show the entropy of entanglement. In them, the region where the entropy reaches its highest values ($S_{\varepsilon}>1.01$) is shown in dark grey. As with the $\Xi$ configuration, it is worth noting that this region gets larger as the cooperation number increases.

The third graphic show a contour plot of the fidelity between neighbouring coherent states. In this, the region where the fidelity drops ($F<0.97$ is emphasised although fidelity drops to values near zero) is shown in dark grey. Irregularities appear due to numerical errors in the energy surface's minimisation process near the transition.

\begin{figure}
\centering
\includegraphics[scale=0.29]{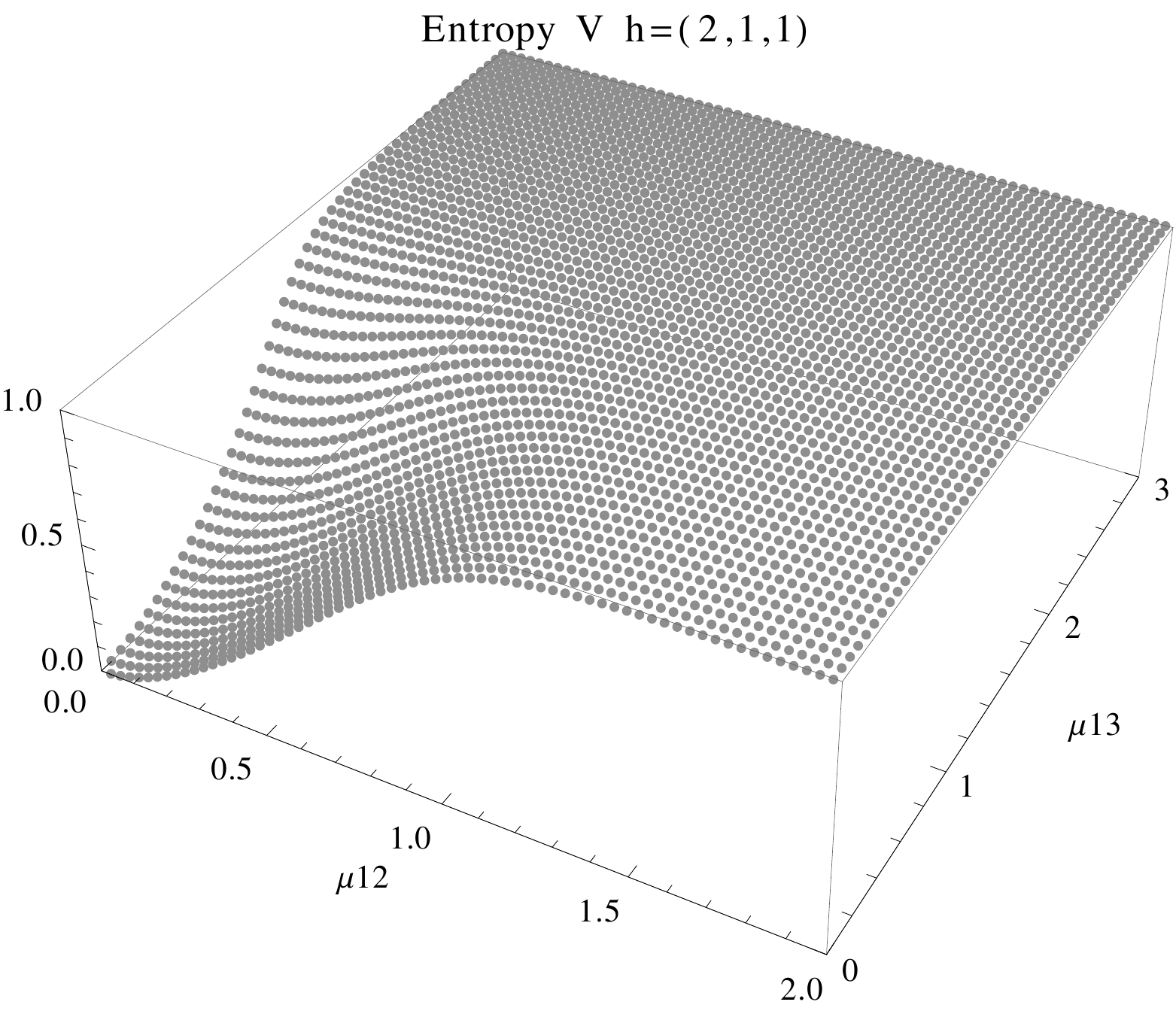}\quad{}\quad{}\includegraphics[scale=0.29]{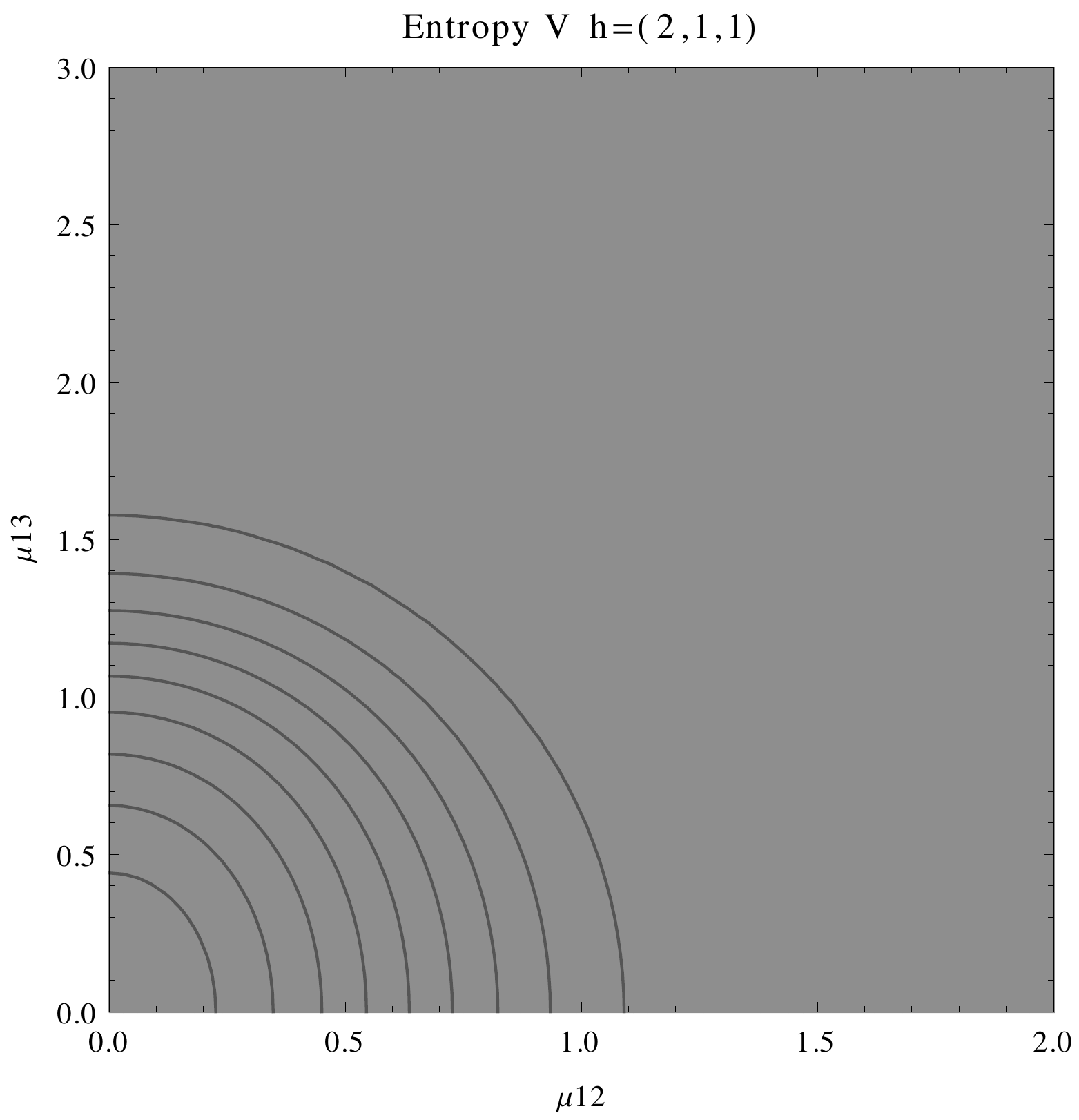}\quad{}\quad{}\includegraphics[scale=0.18]{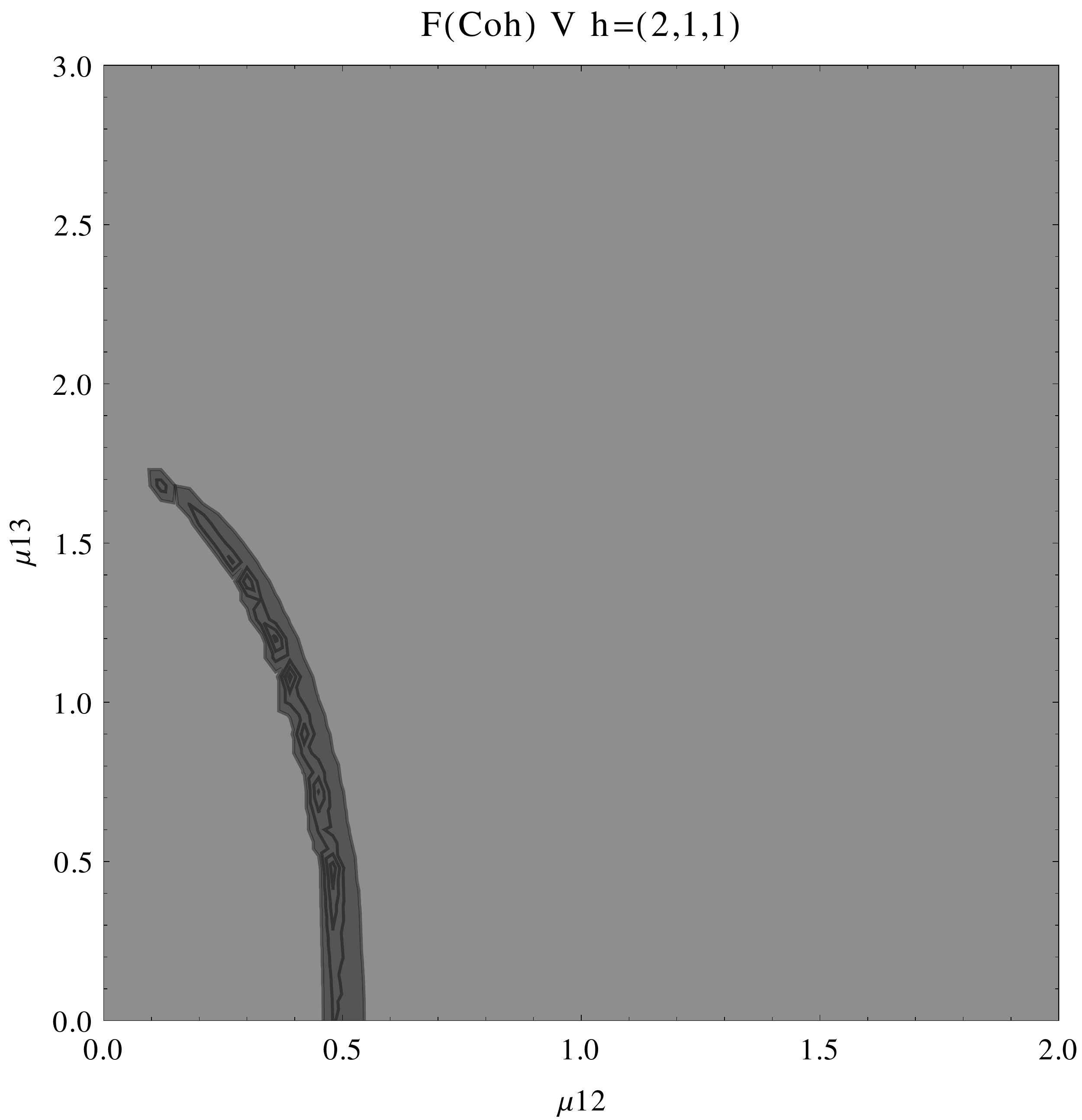}
\caption{(\textbf{Left}) 3D plot of the entropy of entanglement as a function of the coupling parameters $\mu_{12}$ and $\mu_{13}$, the maximum value of the entropy is $S_{\varepsilon}=1$. (\textbf{Center}) Contour plot of the entropy of entanglement as a function of the coupling parameters $\mu_{12}$ and $\mu_{13}$. (\textbf{Right}) Fidelity between neighbouring coherent states as a function of the coupling parameters $\mu_{12}$ and $\mu_{13}$, dark grey region shows the fidelity’s minimum (i.e. the phase transition). All figures use $\omega_{1}=1.\bar{3}$, $\omega_{2}=1.\bar{6}$, $\Omega=0.5$ and correspond to the $V$ configuration and the $h=(2,1,1)$ representation.}\label{fig:10}
\end{figure}

\begin{figure}
\centering
\includegraphics[scale=0.29]{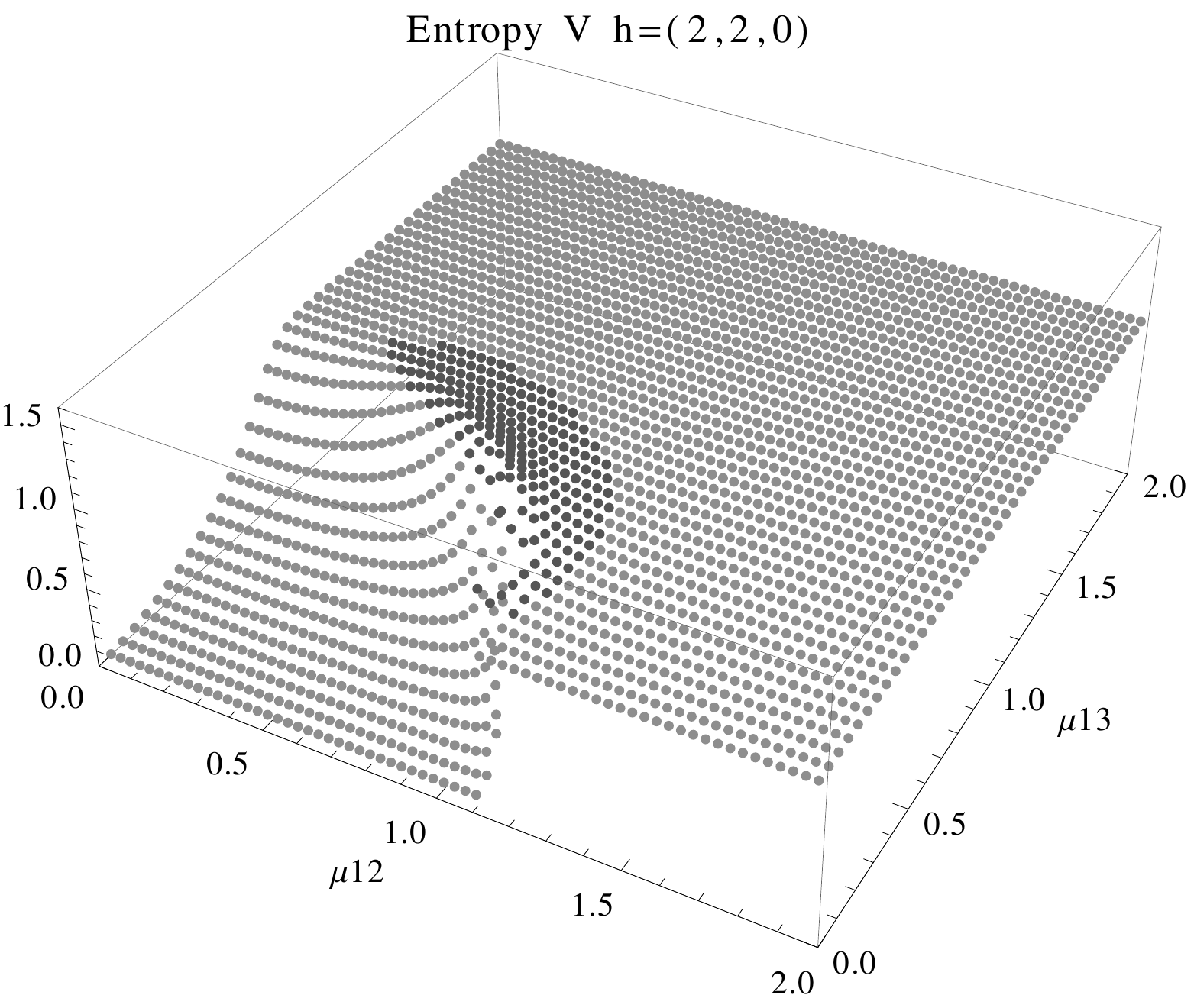}\quad{}\quad{}\includegraphics[scale=0.29]{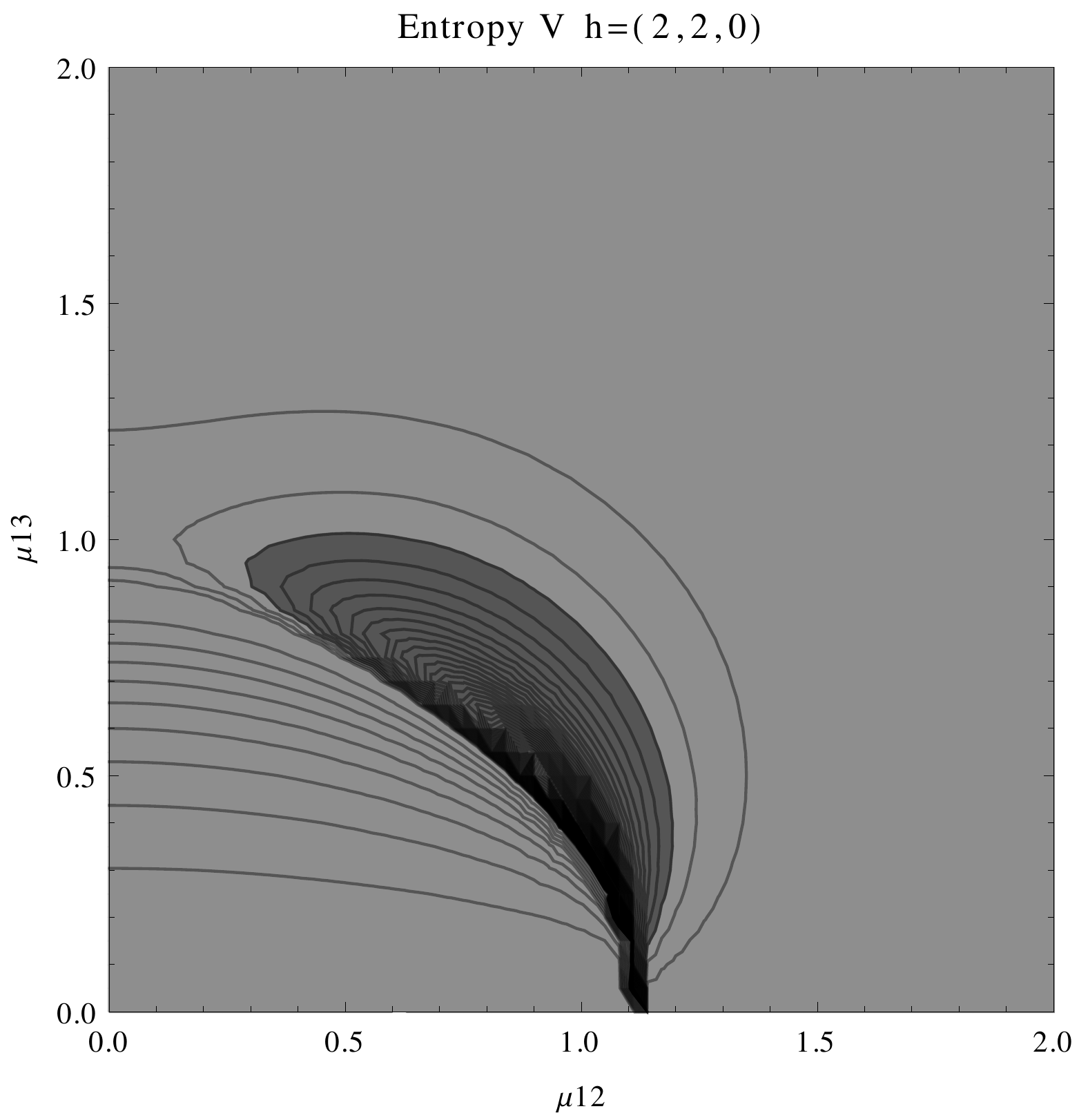}\quad{}\quad{}\includegraphics[scale=0.18]{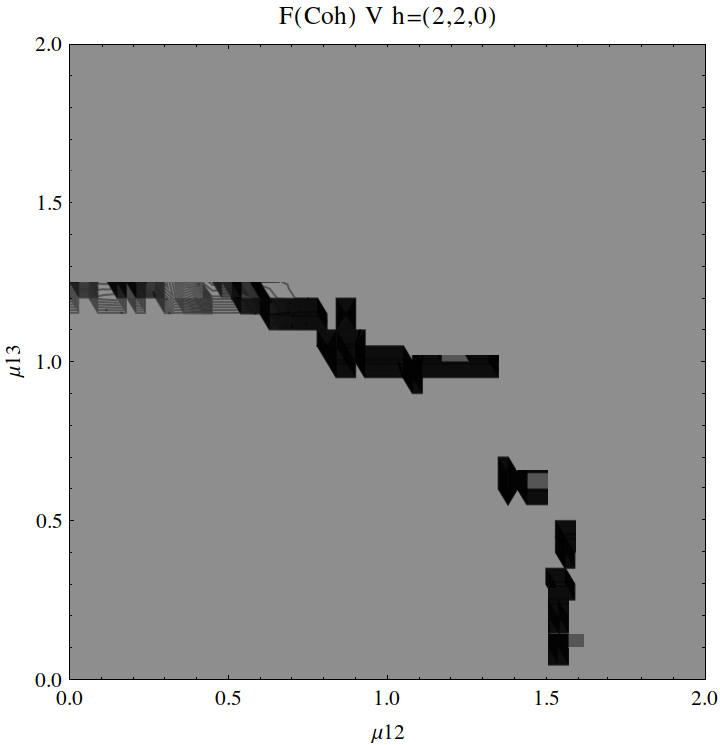}
\caption{(\textbf{Left}) 3D plot of the entropy of entanglement as a function of the coupling parameters $\mu_{12}$ and $\mu_{13}$, the maximum value of the entropy is $S_{\varepsilon}=1.55$ and the region where $S_{\varepsilon}>1.03$ is shown in dark grey. (\textbf{Center}) Contour plot of the entropy of entanglement as a function of the coupling parameters $\mu_{12}$ and $\mu_{13}$, the region where $S_{\varepsilon}>1.03$ is shown in dark grey. (\textbf{Right}) Fidelity between neighbouring coherent states as a function of the coupling parameters $\mu_{12}$ and $\mu_{13}$, dark grey region shows the fidelity’s minimum (i.e. the phase transition). All figures use $\omega_{1}=1.\bar{3}$, $\omega_{2}=1.\bar{6}$, $\Omega=0.5$ and correspond to the $V$ configuration and the $h=(2,2,0)$ representation.}\label{fig:11}
\end{figure}

\begin{figure}
\centering
\includegraphics[scale=0.29]{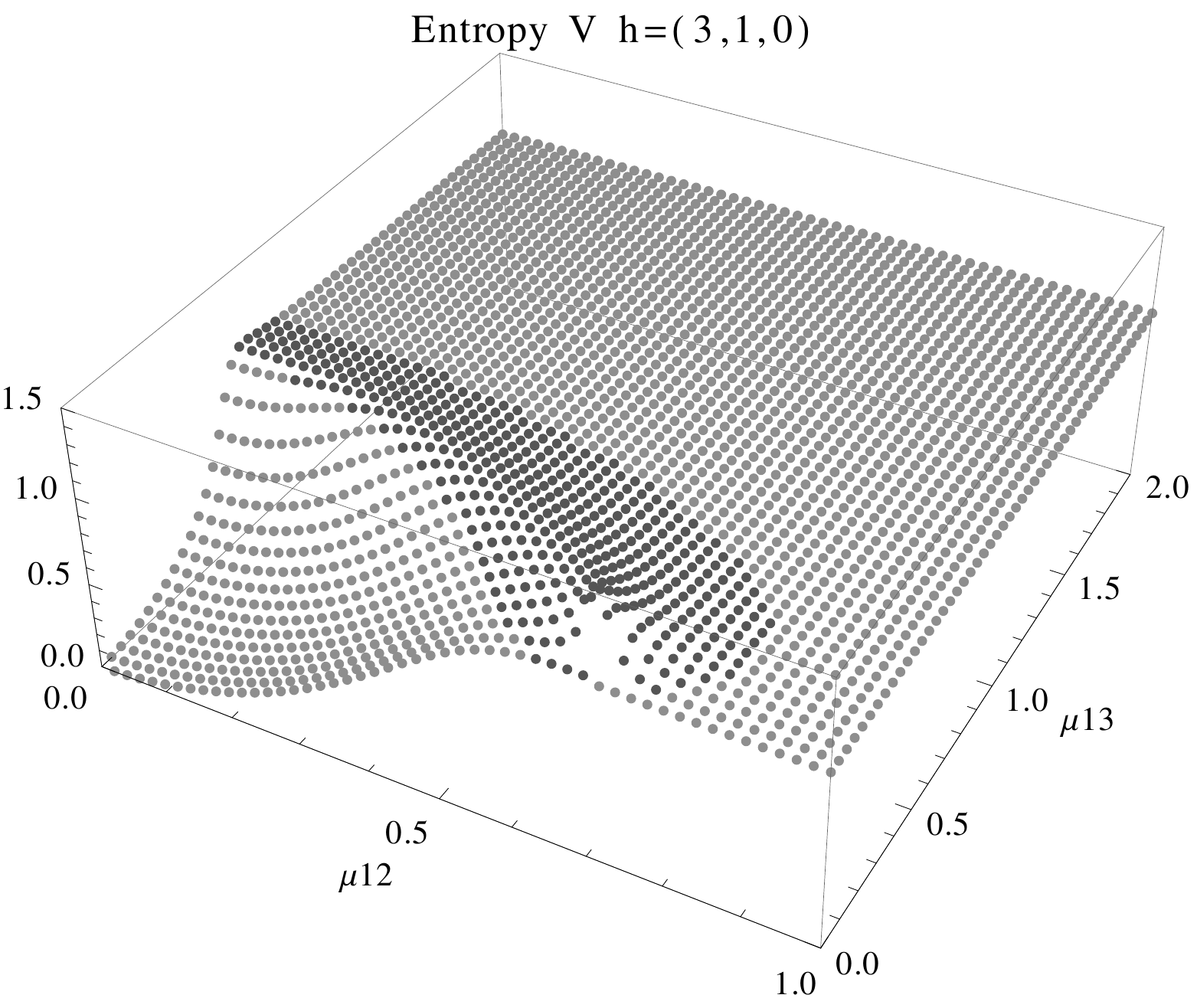}\quad{}\quad{}\includegraphics[scale=0.29]{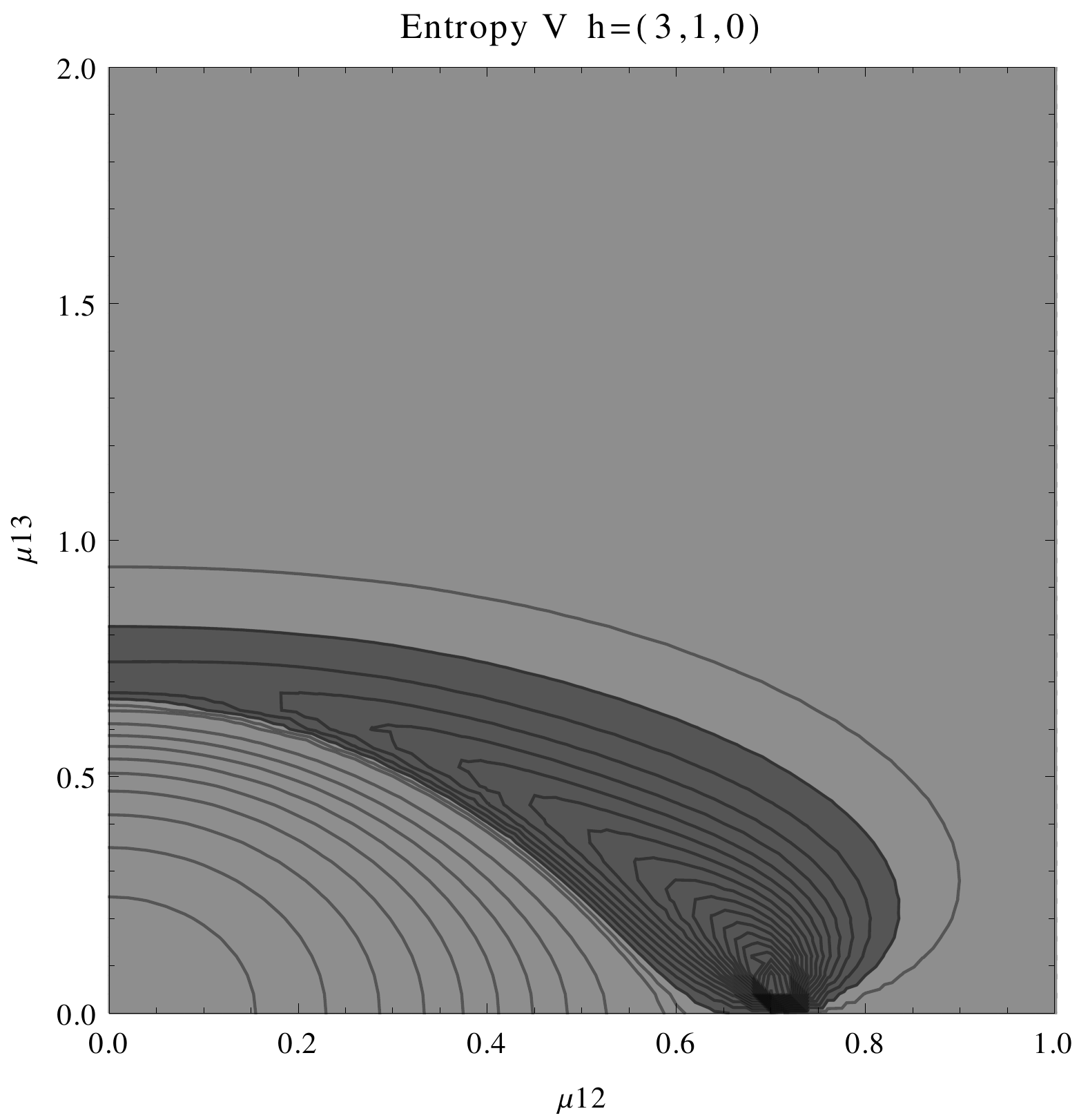}\quad{}\quad{}\includegraphics[scale=0.18]{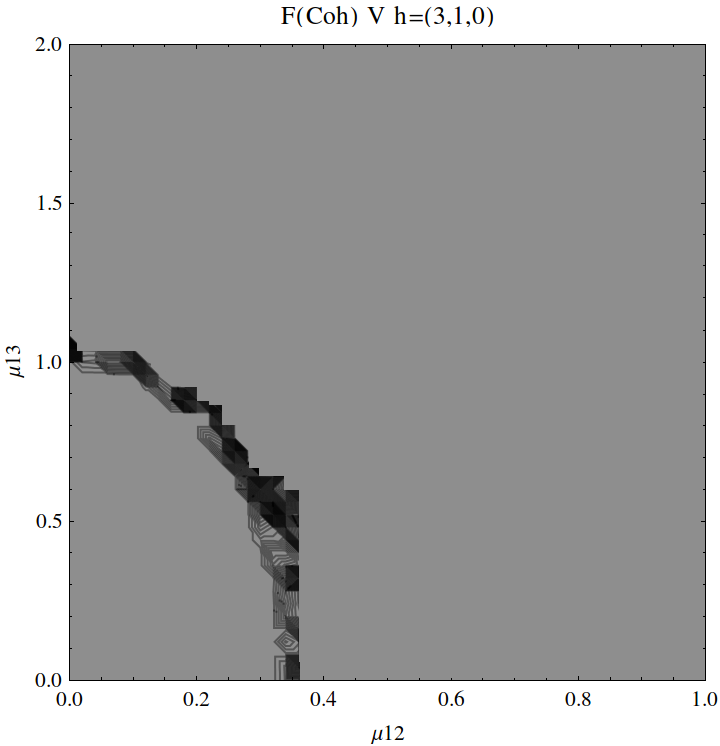}
\caption{(\textbf{Left}) 3D plot of the entropy of entanglement as a function of the coupling parameters $\mu_{12}$ and $\mu_{13}$, the maximum value of the entropy is $S_{\varepsilon}=1.4$ and the region where $S_{\varepsilon}>1.03$ is shown in dark grey. (\textbf{Center}) Contour plot of the entropy of entanglement as a function of the coupling parameters $\mu_{12}$ and $\mu_{13}$, the region where $S_{\varepsilon}>1.03$ is shown in dark grey. (\textbf{Right}) Fidelity between neighbouring coherent states as a function of the coupling parameters $\mu_{12}$ and $\mu_{13}$, dark grey region shows the fidelity’s minimum (i.e. the phase transition). All figures use $\omega_{1}=1.\bar{3}$, $\omega_{2}=1.\bar{6}$, $\Omega=0.5$ and correspond to the $V$ configuration and the $h=(3,1,0)$ representation.}\label{fig:12}
\end{figure}

\begin{figure}
\centering
\includegraphics[scale=0.29]{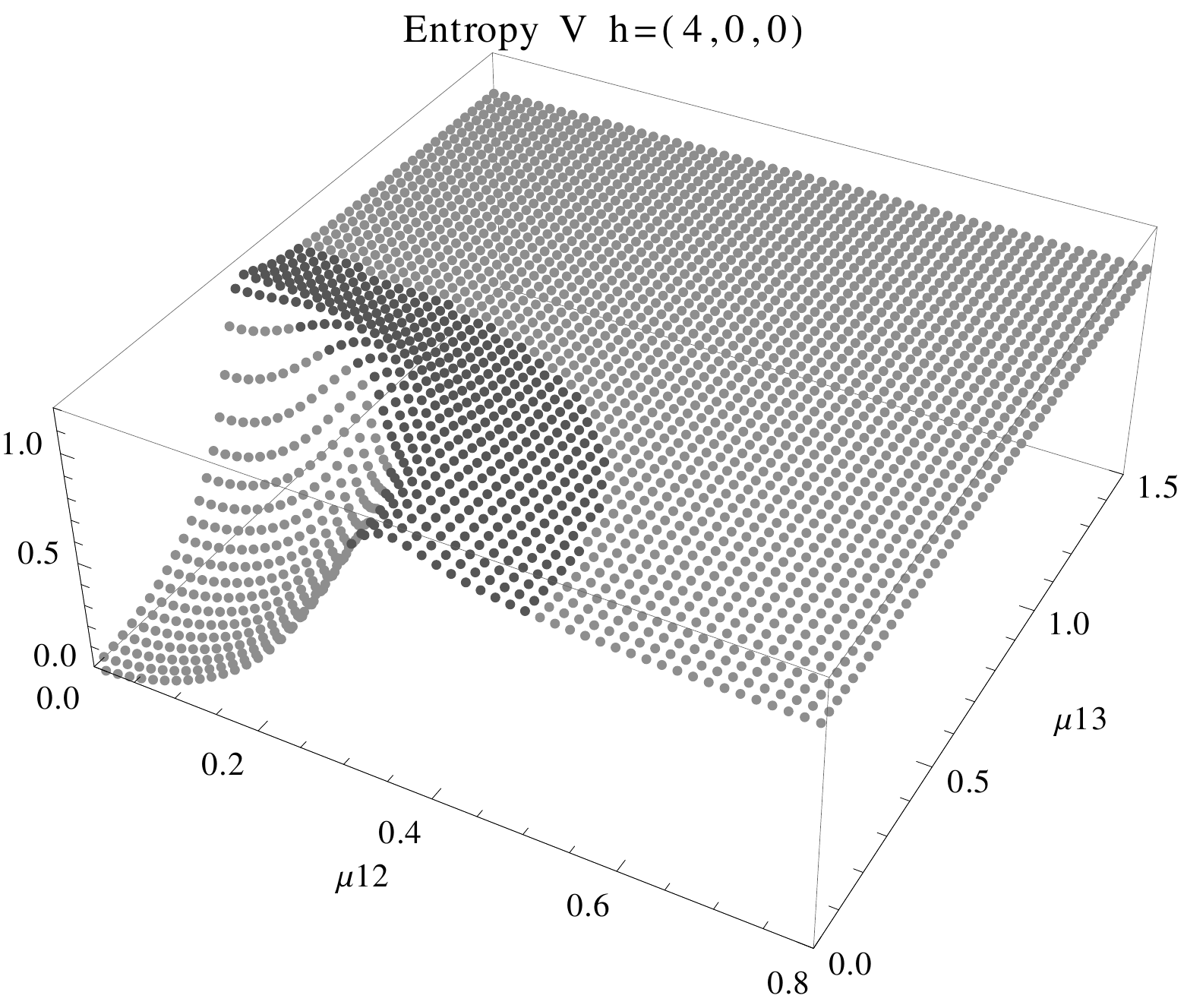}\quad{}\quad{}\includegraphics[scale=0.29]{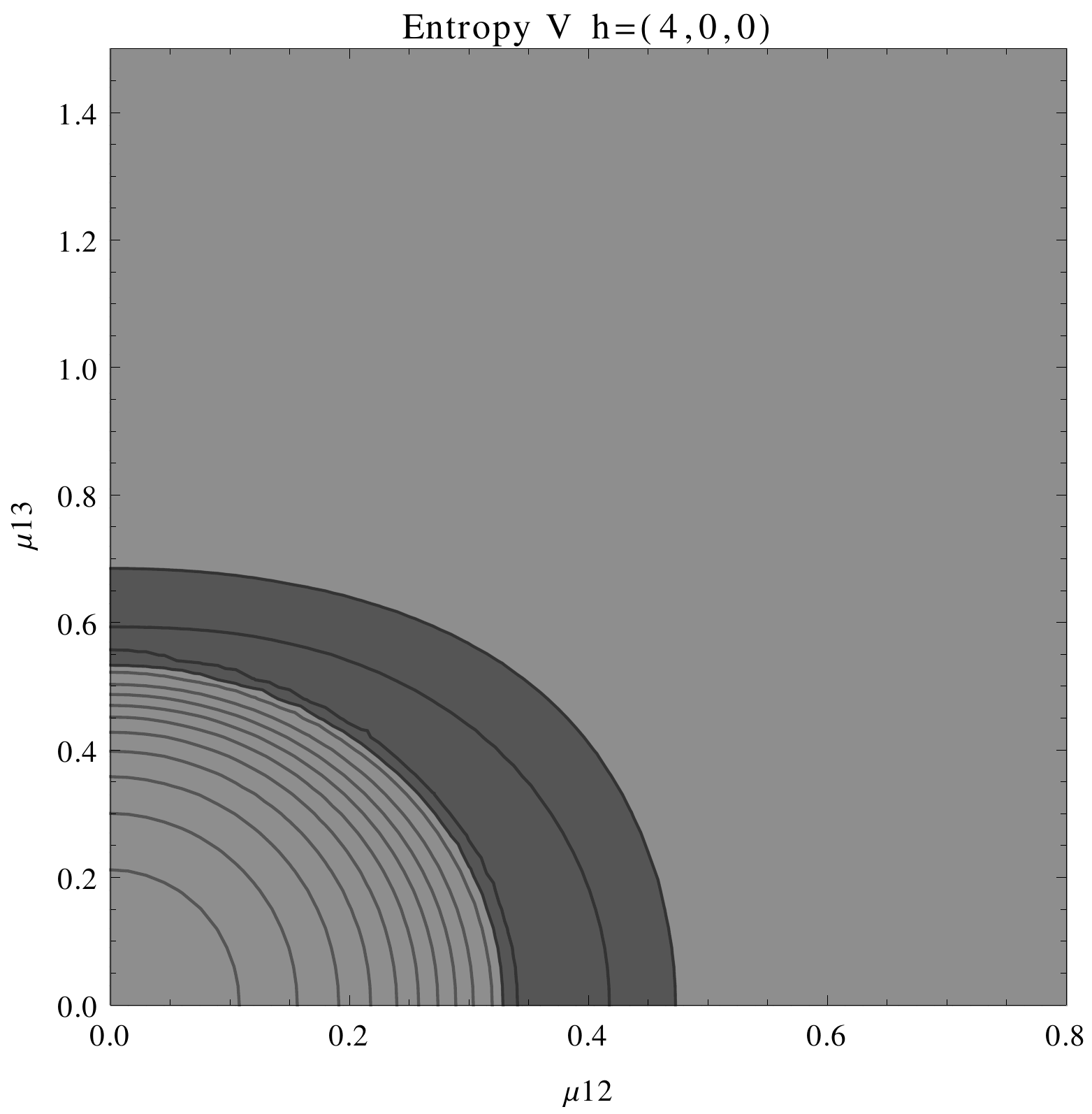}\quad{}\quad{}\includegraphics[scale=0.18]{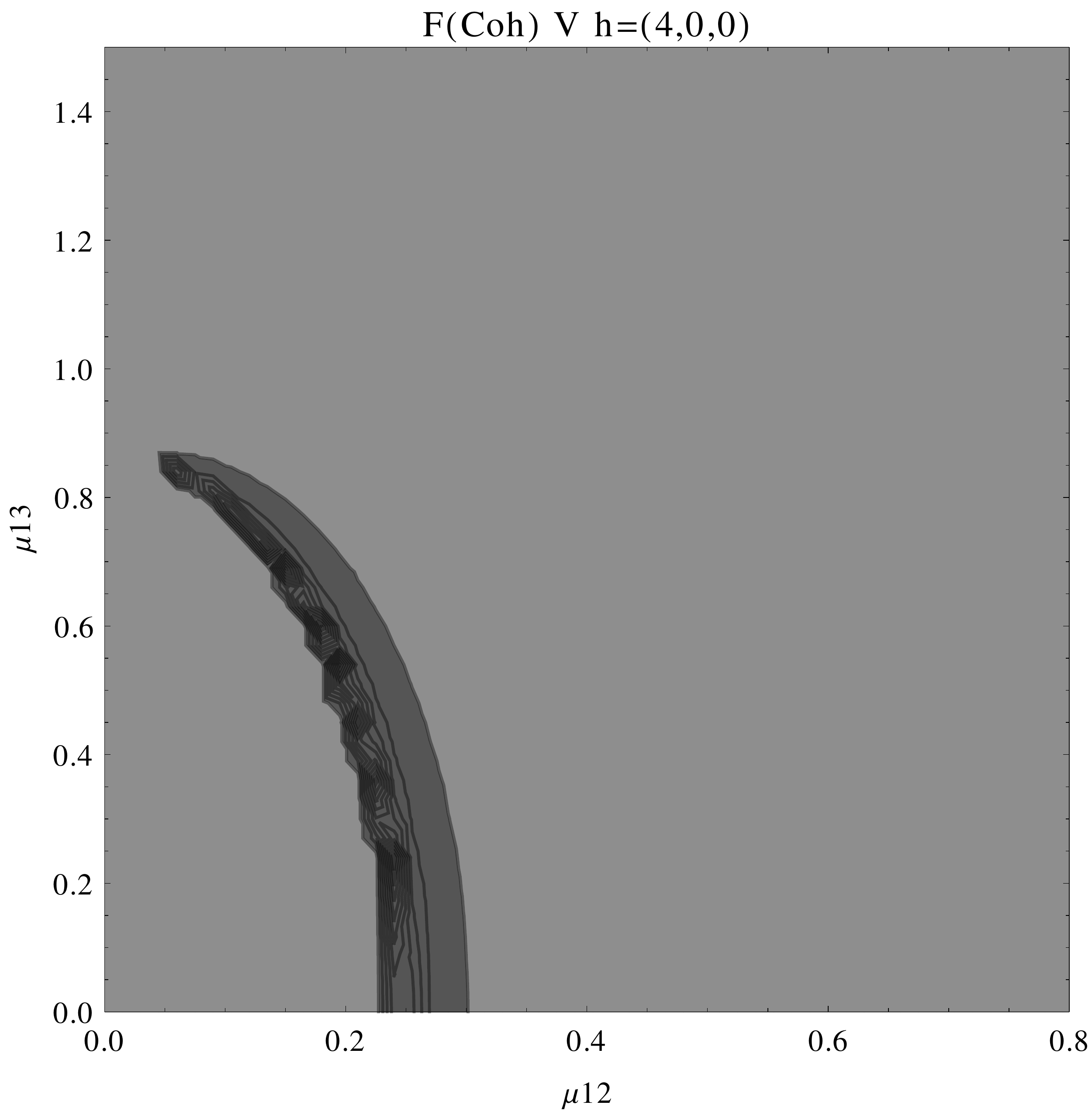}
\caption{(\textbf{Left}) 3D plot of the entropy of entanglement as a function of the coupling parameters $\mu_{12}$ and $\mu_{13}$, the maximum value of the entropy is $S_{\varepsilon}=1.15$ and the region where $S_{\varepsilon}>1.03$ is shown in dark grey. (\textbf{Center}) Contour plot of the entropy of entanglement as a function of the coupling parameters $\mu_{12}$ and $\mu_{13}$, the region where $S_{\varepsilon}>1.03$ is shown in dark grey. (\textbf{Right}) Fidelity between neighbouring coherent states as a function of the coupling parameters $\mu_{12}$ and $\mu_{13}$, dark grey region shows the fidelity’s minimum (i.e. the phase transition). All figures use $\omega_{1}=1.\bar{3}$, $\omega_{2}=1.\bar{6}$, $\Omega=0.5$ and correspond to the $V$ configuration and the $h=(4,0,0)$ representation.}\label{fig:13}
\end{figure}

Finally, we present the results for atoms in the $V$ configuration in figures \ref{fig:10} to \ref{fig:13}, for all four possible cooperation numbers. The first two graphics (from left to right) show the entropy of entanglement, in them, the region where the entropy reaches its highest values ($S_{\varepsilon}>1.03$) is shown in dark grey.  As in the previous configurations, it's worth noting that this region gets larger as the cooperation number increases.

The third graphic show a contour plot of the fidelity between neighbouring coherent states, in this, the region where the fidelity drops ($F<0.97$ is emphasised although fidelity drops to values near zero) is shown in dark grey. Irregularities appear due to numerical errors in the energy surface's minimisation process near the transition.

\section{Discussion and Conclusions}

Figures \ref{fig:2} to \ref{fig:5} show the results obtained for the $\Xi$ configuration, these suggest the existence of at least two quantum phases at zero temperature for all representations and cooperation numbers; these are the so-called {\it normal} and {\it collective} regions. Although it has been already shown that this configuration has a triple point (i.e. three phases) in the symmetric representation \cite{key-28}, the discrepancy leads us to conclude that the entropy of entanglement is just sensitive to the transition between normal and super-radiant phases but not between possible transitions within these regions.

In the $\Lambda$ and $V$ configurations there is evidence of only two phases in the phase space of its ground state at zero temperature, the normal and collective regions, and these are well determined by the entanglement entropy.

An interesting pattern present in the three configurations is that of the increase in the sensitivity of the entropy of entanglement as the cooperation number tends to the actual number of atoms. This can be seen by noting that the region where the entropy reaches its highest values gets larger as the cooperation number increases.

It is worthwhile noticing that, while the trial state is a tensor product of coherent states and therefore shows no entanglement between matter and the radiation field, the phase diagrams obtained via these variational states is well displayed by the entropy of entanglement calculated through quantum means. In contrast the latter does not dictate the exact quantum phase transitions for finite $N$~\cite{key-31}; they coincide only in the thermodynamic limit.

From the figures presented, and based on the fact that the coherent QPT and the ``real'' QPT coincide in the thermodynamic limit, we are able to conclude that there is indeed a resemblance between the QPT of the studied system and the highest values of its entropy of entanglement for a finite number of atoms. This conclusion suggests that there are more than one possible states the system can be in at zero temperature, hence its residual entropy must be different from zero.








\vspace{6pt} 


\acknowledgments{This work was partially supported by DGAPA-UNAM under
project IN101217. L.F.Q. thanks CONACyT-M\'exico for financial support
(Grant \#379975).}


\begin{thebibliography}{999}
\bibitem{key-1}
Dicke, R. H. Coherence in Spontaneous Radiation Processes. {\em Phys. Rev.} {\bf 1954}, {\em 93}, 99, DOI: 10.1103/PhysRev.93.99.
\bibitem{key-2}
Hepp, K. and Lieb, H. On the superradiant phase transition for molecules in a quantized radiation field: the dicke maser model. {\em Ann. Phys.} {\bf 1973}, {\em 76}, 360, DOI: 10.1016/0003-4916(73)90039-0.
\bibitem{key-3}
Hepp, K. and Lieb, H. Equilibrium Statistical Mechanics of Matter Interacting with the Quantized Radiation Field. {\em Phys. Rev. A} {\bf 1973}, {\em 8}, 2517, DOI: 10.1103/PhysRevA.8.2517.
\bibitem{key-4}
Wang, Y. and Hioe, F. Phase Transition in the Dicke Model of Superradiance. {\em Phys. Rev. A} {\bf 1973}, {\em 7}, 831, DOI: 10.1103/PhysRevA.8.2517.
\bibitem{key-5}
Baumann, K., Guerlin, C., Brennecke, F. and Esslinger, T. Dicke quantum phase transition with a superfluid gas in an optical cavity. {\em Nature (London)} {\bf 2010}, {\em 464}, 1301, DOI: 10.1038/nature09009.
\bibitem{key-6}
Nagy, D., Kónya, G., Szirmai, G. and Domokos, P. Dicke-Model Phase Transition in the Quantum Motion of a Bose-Einstein Condensate in an Optical Cavity. {\em Phys. Rev. Lett.} {\bf 2010}, {\em 104}, 130401, DOI: 10.1103/PhysRevLett.104.130401.
\bibitem{key-7}
Brandes, T. Coherent and collective quantum optical effects in mesoscopic systems. {\em Phys. Rep.} {\bf 2005}, {\em 408}, 315, DOI: 10.1016/j.physrep.2004.12.002.
\bibitem{key-8}
Chen, G., Chen, Z. and Liang, J. Simulation of the superradiant quantum phase transition in the superconducting charge qubits inside a cavity. {\em Phys. Rev. A} {\bf 2007}, {\em 76}, 055803, DOI: 10.1103/PhysRevA.76.055803.
\bibitem{key-9}
Lambert, N., Emary, C. and Brandes, T. Entanglement and the Phase Transition in Single-Mode Superradiance. {\em Phys. Rev. Lett.} {\bf 2004}, {\em 92}, 073602, DOI: 10.1103/PhysRevLett.92.073602.
\bibitem{key-10}
Lambert, N., Emary, C. and Brandes, T. Entanglement and entropy in a spin-boson quantum phase transition. {\em Phys. Rev. A} {\bf 2005}, {\em 71}, 053804, DOI: 10.1103/PhysRevA.71.053804.
\bibitem{key-11}
Yoo, H. I. and Eberly, J. H. Dynamical theory of an atom with two or three levels interacting with quantized cavity fields. {\em Phys. Rep.} {\bf 1985}, {\em 118}, 239, DOI: 10.1016/0370-1573(85)90015-8.
\bibitem{key-12}
Civitarese, O. and Reboiro, M. Atomic squeezing in three level atoms. {\em Phys. Lett. A} {\bf 2006}, {\em 357}, 224, DOI: 10.1016/j.physleta.2006.04.043.
\bibitem{key-13}
Abdel-Wahab, N. H. A three-level atom interacting with a single mode cavity field: different configurations. {\em Phys. Scr.} {\bf 2007}, {\em 76}, 244, DOI: 10.1088/0031-8949/76/3/006.
\bibitem{key-14}
Abdel-Wahab, N. H. A three-level atom interacting with a single mode cavity field: double $\Xi$-configuration. {\em Mod. Phys. Lett. B} {\bf 2008}, {\em 22}, 2587, DOI: 10.1142/S0217984908016868.
\bibitem{key-15}
Hayn, M., Emary, C. and Brandes, T. Phase transitions and dark-state physics in two-color superradiance. {\em Phys. Rev. A} {\bf 2011}, {\em 84}, 053856, DOI: 10.1103/PhysRevA.84.053856.
\bibitem{key-16}
Hayn, M., Emary, C. and Brandes, T. Superradiant phase transition in a model of three-level-$\Lambda$ systems interacting with two bosonic modes. {\em Phys. Rev. A} {\bf 2012}, {\em 86}, 063822, DOI: 10.1103/PhysRevA.86.063822.
\bibitem{key-17}
Cordero, S., López–Peña, R., Castaños, O. and Nahmad–Achar, E. Quantum phase transitions of three-level atoms interacting with a one-mode electromagnetic field. {\em Phys. Rev. A} {\bf 2013}, {\em 87}, 023805, DOI: 10.1103/PhysRevA.87.023805.
\bibitem{key-18}
Cordero, S., Castaños, O., López–Peña, R. and Nahmad–Achar, E. A semi-classical versus quantum description of the ground state of three-level atoms interacting with a one-mode electromagnetic field. {\em J. Phys. A: Math. Theor.} {\bf 2013}, {\em 46}, 505302, DOI: 10.1088/1751-8113/46/50/505302.
\bibitem{key-19}
Cordero, S., Nahmad-Achar, E., López-Peña, R. and Castaños, O. Polychromatic phase diagram for n-level atoms interacting with $\ell$ modes of an electromagnetic field. {\em Phys. Rev. A} {\bf 2015}, {\em 92}, 053843, DOI: 10.1103/PhysRevA.92.053843.
\bibitem{key-20}
Cordero, S., Castaños, O., López–Peña, R. and Nahmad–Achar, E. Variational study of $\lambda$ and N atomic configurations interacting with an electromagnetic field of two modes. {\em Phys. Rev. A} {\bf 2016}, {\em 94}, 013802, DOI: 10.1103/PhysRevA.94.013802.
\bibitem{key-21}
Kozhekin, A. E., Mølmer, K. and Polzik, E. Quantum memory for light. {\em Phys. Rev. A} {\bf 2000}, {\em 62}, 033809, DOI: 10.1103/PhysRevA.62.033809.
\bibitem{key-22}
Gorshkov, A., André, A., Fleischhauer, M., Sørensen, A. and Lukin, M. Universal Approach to Optimal Photon Storage in Atomic Media. {\em Phys. Rev. Lett.} {\bf 2007}, {\em 98}, 123601, DOI: 10.1103/PhysRevLett.98.123601.
\bibitem{key-23}
Nunn, J., Walmsley, I. A.,  Raymer, M. G., Surmacz, K., Waldermann, F. C., Wang, Z. and Jaksch, D. Mapping broadband single-photon wave packets into an atomic memory. {\em Phys. Rev. A} {\bf 2007}, {\em 75}, 011401(R), DOI: 10.1103/PhysRevA.75.011401.
\bibitem{key-24}
Morton, J. L., Tyryshkin, A. M., Brown, R. M., Shankar, S., Lovett, B. W., Ardavan, A., Schenkel, T., Haller, E. E., Ager, J. W. and Lyon, S. A. Solid-state quantum memory using the 31P nuclear spin. {\em Nature} {\bf 2008}, {\em 455}, 1085, DOI: 10.1038/nature07295.
\bibitem{key-25}
Quezada, L. F. and Nahmad-Achar, E. Characterization of the quantum phase transition in a two-mode Dicke model for different cooperation numbers. {\em Phys. Rev. A} {\bf 2017}, {\em 95}, 013849, DOI: 10.1103/PhysRevA.95.013849.
\bibitem{key-26}
Zanardi, P. and Paunković, N. Ground state overlap and quantum phase transitions. {\em Phys. Rev. E} {\bf 2006}, {\em 74}, 031123, DOI: 10.1103/PhysRevE.74.031123.
\bibitem{key-27}
Castaños, O., Nahmad-Achar, E., López-Peña, R., and Hirsch, J. G. Universal critical behavior in the Dicke model. {\em Phys. Rev. A} {\bf 2012}, {\em 86}, 023814, DOI: 10.1103/PhysRevA.86.023814.
\bibitem{key-28}
Nahmad-Achar, E. Cordero, S., López-Peña, R. and Castaños, O. A triple point in 3-level systems. {\em J. Phys. A: Math. Theor.} {\bf 2014}, {\em 47}, 455301, DOI: 10.1088/1751-8113/47/45/455301.
\bibitem{key-29}
Gelfand, I. M. and Tsetlin, M. L. Finite-dimensional representations of the group of unimodular matrices. {\em Dokl.Akad.Nauk} {\bf 1950}, {\em 71}, 825.
\bibitem{key-30}
Nahmad-Achar, E., Cordero, S., Castaños, O. and López-Peña, R. Phase diagrams of systems of two and three levels in the presence of a radiation field. {\em Phys. Scr.} {\bf 2015}, {\em 90}, 7, DOI: 10.1088/0031-8949/90/7/074026.
Reference book
\bibitem{key-31}
Quezada, L. F. {\em Cooperativity in Matter-Radiation Interaction Models}; Ph. D. Thesis, National University of Mexico (UNAM), Mexico, Cd. Mx., 2018.
\end{thebibliography}
\end{document}